\documentclass[%
 notitlepage,
superscriptaddress,
 amsmath,amssymb,
 aps,
 reprint,
prl,
nofootinbib
]{revtex4-1}
\usepackage[dvipsnames]{xcolor}

\usepackage{graphicx}
\usepackage{dcolumn}
\usepackage{bm}
\usepackage{hyperref}
\hypersetup{colorlinks = true, linkcolor = [rgb]{0.19411,0.51882,0.667058}, urlcolor = [rgb]{0.125490,0.29542,0.1647058}, citecolor = [rgb]{0.75882,0.37411,0.14117}}

\usepackage{mathtools}
\usepackage{setspace}

\usepackage{amssymb}

\usepackage{braket}

\graphicspath{{../images/}}
\usepackage[toc,page]{appendix}

\newcommand{\nn}{\nonumber \\}

\usepackage{braket}
\DeclareMathOperator{\tr}{Tr}

\def\>{\rangle}
\def\<{\langle}

\def\section#1{{\par\em #1:--- }}
\usepackage[bottom]{footmisc}

\usepackage[normalem]{ulem}
\begin{document}

\title{Imaging stars with quantum error correction }

\author{Zixin Huang}
\email{zixin.huang@mq.edu.au}
\affiliation{Centre for Engineered Quantum Systems, School of Mathematical and Physical Sciences, Macquarie University, NSW 2109, Australia}

\author{Gavin K. Brennen}
\email{gavin.brennen@mq.edu.au}
\affiliation{Centre for Engineered Quantum Systems, School of Mathematical and Physical Sciences, Macquarie University, NSW 2109, Australia}

\author{Yingkai Ouyang}
\email{oyingkai@gmail.com}
\affiliation{Department of Electrical and Computer Engineering, National University of Singapore, Singapore}
\affiliation{Centre of Quantum Technologies, National University of Singapore, Singapore}

\begin{abstract}\noindent
The development of high-resolution, large-baseline optical interferometers would revolutionize astronomical imaging. However, classical techniques are hindered by physical limitations including loss, noise, and the fact that the received light is generally quantum in nature. We show how to overcome these issues using quantum communication techniques.
We present a general framework for using quantum error correction codes for protecting and imaging starlight received at distant telescope sites. 
In our scheme, the quantum state of light is coherently captured into a non-radiative atomic state via Stimulated Raman Adiabatic Passage, which is then imprinted into a quantum error correction code. 
The code protects the signal during subsequent potentially noisy operations necessary to extract the image parameters. We show that even a small quantum error correction code can offer significant protection against noise. 
For large codes, we find noise thresholds below which the information can be preserved.
Our scheme represents an application for near-term quantum devices that can increase imaging resolution beyond what is feasible using classical techniques.

\end{abstract}
\date{\today}
 
\maketitle

The performance of an imaging system is limited by diffraction: the resolution is proportional to its aperture and inversely proportional to the wavelength $\lambda$.  Together these place a fundamental limit on how well one can image the objects of interest. Typical techniques employed to enable quantum sensing and quantum imaging to surpass classical limits utilise entanglement \cite{giovannetti2006quantum,giovannetti2011advances}, source engineering \cite{lichtman2005fluorescence}, or squeezing \cite{casacio2021quantum} to suppress intensity fluctuations. 
These techniques require manipulating the objects or illuminating them with light that has special properties. However, often it is the case, e.g. for astronomy, we cannot illuminate the objects of interest. Rather all we can do is analyse the light that reaches us.

Several challenges hinder the progress in building large-baseline optical interferometers, one of which is the presence of noise and transmission loss that ultimately limits the distance between telescope sites. Quantum technologies can help bypass transmission losses: using quantum memories and entanglement, we can replace the direct optical links, allowing us to go to large distances.
In the most direct approach, \cite{PhysRevLett.123.070504}, we could store the signal into atomic states and perform operations to extract the information; however, such states are sensitive to optical decay and other decoherences.
One way to combat noise is to employ quantum error correction (QEC). QEC has been predominantly studied in the context of quantum computation \cite{RevModPhys.87.307} and specialised sensing protocols \cite{zhou2018achieving,PhysRevX.7.041009,PhysRevA.94.012101,shettell2021practical,PhysRevLett.112.150802,PhysRevLett.112.080801,PhysRevLett.116.230502,PhysRevLett.122.040502,ouyang2019robust,ouyang2020weight}.

In this paper, we produce a general framework for using QEC codes to protect the information in the received light. This is the first time that QEC has been applied to a quantum parameter estimation task where the probe state need not be prepared by the experimenter. 
We eliminate optical decay in quantum memories by coupling light into non-radiative atomic qubit states via the well-developed process known as STImulated Raman Adiabatic Passage (STIRAP) \cite{RevModPhys.89.015006}. Our scheme can also accommodate for multi-photon events.
{
Two results \cite{PhysRevLett.109.070503,PhysRevLett.123.070504} have investigated the potential for large-baseline entanglement-aided imaging;
 here we explicitly take into account noise sources and propose a robust encoding of the signal into quantum memories.}
\begin{figure*}[t]
\includegraphics[trim = 0cm 0.0cm 0cm 0cm, clip, width=1.0\linewidth]{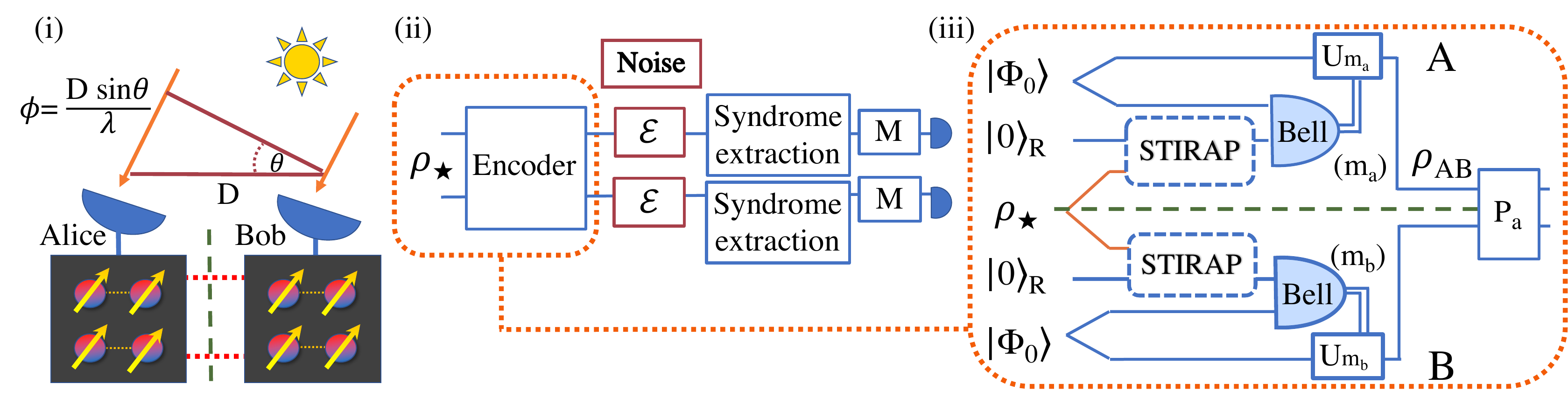} 
 \caption{\label{f:memoryscheme} \textbf{Overview of our protocol. }(i) Light at wavelength $\lambda$  is collected at two sites (Alice and Bob) separated by distance $D$, each holding quantum memories and sharing Bell pairs. (ii) A general framework for the process of encoding, error correction, and measuring the signal. The starlight $\rho_\star$ is input into the encoder, which outputs a logical state of a quantum code. The memory qubits and subsequent operations are potentially noisy as modelled by local channels $\mathcal{E}$. 
 Any correctible error is detected by syndrome measurements and the final step is a local Clifford measurement ${\mathrm M}$, which depends on the error syndromes, and extracts the parameters of interest from the state.
  (iii) A detailed schematic showing how to encode the starlight into a logical state. The green dashed line denotes the spatial separation between Alice and Bob. ${\rm U_m}$ are Pauli corrections depending on the outcome of local Bell measurements. ${\mathrm P_a}$ is a parity measurement to project out the vacuum component (and multiphoton contributions) and is the only part that uses non-local resources.}
\end{figure*}
Any imaging task can be translated into a parameter estimation task, where the quantity of interest is the quantum Fisher information (QFI). 
We show that even a small QEC code can offer significant protection against noise which degrades resolution.

\section{The protocol}
\label{sec:protocol}
We show an overview of our protocol in Fig.~\ref{f:memoryscheme}. Consider a two-site scenario (Alice and Bob), each holding a telescope station and are separated by large distances.
The layer of quantum technology is schematically shown in panel (i), where light from astronomical sources is collected: Alice and Bob share pre-distributed entanglement, and the sites each contain quantum memories into which the light is captured.  This becomes the encoder operation in (ii): they each prepare (locally) their set of qubits into some QEC code. The received state $\rho_{\star}$ is imprinted onto the code via an encoder, resulting in the logical state $\rho_{AB}$ shared between Alice and Bob. The state is thus protected from subsequent noisy operations.
Panel (iii) shows the circuit of the encoder. 
 \color{black}

In the ``encoder" stage, we need to capture the signal into the quantum memories, which involves a light-matter interaction Hamiltonian. In the naive approach, we could use two-level atoms with ground-excited state encoding, $\ket{g}, \ket{e}$, where the energy difference corresponds to that of the photon. If we were to place such an ensemble of excited atoms in a cavity, the atoms will undergo optical decay that can introduce errors or take the state outside the codespace. To circumvent this, we use STIRAP which allows us to coherently couple the incoming light into a non-radiative state of an atom. Unlike the naive approach, we do not need to match the frequency of the signal with the atomic transition, providing more bandwidth and flexibility. The state is then imprinted onto a QEC code.

\section{The model}
\label{sec:model}
We model the incoming signal as a weak thermal state \cite{mandel1995optical} that has been multiplexed into frequency bands narrow enough for interferometry. 
First, consider the case where at most a single photon arrives on the two sites ($\epsilon \ll1$). For higher photon numbers see Appendix. We can describe the optical state by the density matrix
\begin{align}\label{eq:coupled}
\rho_{\star} \approx &(1-\epsilon)\ket{\text{vac},\text{vac}}\bra{\text{vac},\text{vac}}_{A B} +  \nn
&\epsilon \left(\frac{1+\gamma}{2}\right) \ket{\psi_+^\phi}\bra{\psi_+^\phi} + 
\epsilon \left(\frac{1-\gamma}{2}\right)\ket{\psi_-^\phi}\bra{\psi_-^\phi} 
\end{align}
\noindent where $\ket{\psi_\pm^\phi} = (\ket{1_p}_{A} \ket{\text{vac}}_{B} \pm e^{i\phi} \ket{\text{vac}}_{A} \ket{1_p}_{B})/\sqrt{2} $. Here the subscript $p$ denotes a photon Fock state of the corresponding photon number.

The parameters of interest are $\phi$ and $\gamma$, where $\phi \in [0 ,2\pi )$ is related to the location of the sources, 
and $\gamma \in [0,1]$ is proportional to the Fourier transform of the intensity distribution via the van Cittert-Zernike theorem \cite{mandel1995optical}. Optimally estimating $\phi $ and $\gamma$ provides complete information of the source distribution \cite{pearce2017optimal}. See Appendix for a review of the QFI (\cite{caves,caves1,paris2009quantum,PhysRevA.94.052108,PhysRevLett.124.080503,PhysRevA.63.053804,PhysRevLett.85.5098,PhysRevA.95.053837,PhysRevLett.110.240405}). 
Here we consider a two-mode state for simplicity; this easily extends to multi-mode, broadband operation by incorporating the time and frequency-multiplexed encoding~\cite{PhysRevLett.123.070504,PhysRevA.100.022316}. Our entanglement cost is the same as those in Refs.~\cite{PhysRevLett.123.070504,PhysRevA.100.022316} if we use their efficient time-bin encoding.
 Memoryless schemes similar to Ref.~\cite{PhysRevLett.109.070503} have an entanglement cost scaling as $1/\epsilon$, whereas ours and those of Refs.~\cite{PhysRevLett.123.070504,PhysRevA.100.022316} scale as 
$\log(1/\epsilon)$.  

The novel steps introduced in this paper include 1) transferring the stellar photon modes to atomic states via the STIRAP method, 2) the Bell measurement that encodes the stellar photon into a QEC code, and 3) the quantitative advantage afforded by QEC. 
\color{black}

\section{Stimulated Raman Adiabatic Passage}
\label{sec:stirap}

\begin{figure}[h]
\includegraphics[trim = 0cm 0.0cm 0cm 0cm, clip, width=1.0\linewidth]{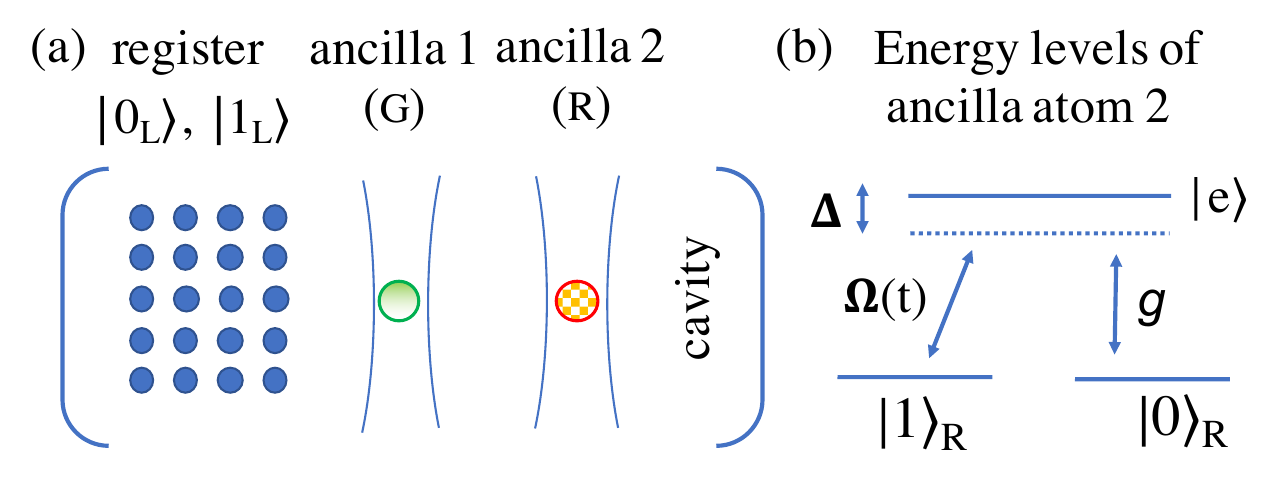} 
 \caption{\label{f:stirap} \textbf{Cavity-assisted coherent single-photon transfer.} (a) A system of qubits has logical states $\ket{0_L},\ket{1_L}$; ancillary qubit 1 is initially prepared into a Bell state with the register, $1/\sqrt2 (\ket{0_L 0_G} + \ket{1_L 1_G})$ ; ancilla 2 is used in the STIRAP interaction to interact with the star photon. (b) Energy levels the ancilla atom 2 used for the STIRAP interaction.}
\end{figure}

STIRAP is inherently robust to parameter errors and resilient to certain types of noise, emerging as a popular tool in quantum information. It is immune to loss through spontaneous emission, and robust against small variations of experimental conditions, e.g. laser intensity and pulse timing \cite{RevModPhys.89.015006}. 
It allows the complete transfer of population along a three-level chain $\ket{0}_R\rightarrow\ket{e}\rightarrow\ket{1}_R$ (Fig.~\ref{f:stirap} (b)) from an initially populated state $\ket{0_R}$ to a target state $\ket{1}_R$ via a light-matter interaction Hamiltonian.
We depict our set-up in Fig.~\ref{f:stirap}: (a) inside a cavity, we use three systems. We denote the blue array as the register.
The blue array is initialised in a codespace of a QEC code encoding a single logical qubit, spanned by its logical codewords $\ket{0_L}$ and $\ket{1_L}$. We also need ancilla qubit 1 (green atom with gradient fill), and ancilla atom 2 
(red atom with check pattern). Note the three types of matter qubits could consist of different electronic sublevels of the same species of atom if desired. 

Panel (b) depicts the energy levels of ancilla atom 2. This atom has three energy levels: the excited state $\ket{e}$, and two ground states, $\ket{0}_R$ and $~\ket{1}_R$ that we have assumed degenerate, though this is not essential. 
The energy difference between $\ket{0}_R$ and $\ket{e}$ is $\omega_0$. 
Ancilla atom 2 is optically trapped. The cavity coupling between $\ket{0}_R$ and $\ket{e}$ is denoted $g$, which is a fixed parameter that depends on the properties of the atom and the cavity;
the time-dependent Rabi frequency on the transition $\ket{1}_R$ to $\ket{e}$ is denoted $\Omega(t)$. The parameter $\Omega(t)$ is tunable via changing the intensity of another laser, which has frequency $\omega_L$, and has detuning $\Delta$ from $\omega_0$.

Defining $n$ to be the number of photons in the cavity, the STIRAP Hamiltonian is
\begin{align}
H_\text{stirap}(t) = &~ \omega_0 \ket{e}\bra{e} + \omega_c a^\dagger a  + \nn
                 & \Omega(t) e^{-i \omega_L t} \ket{e}\bra{1}_R + \Omega(t)^*e^{i \omega_L t}\ket{1}_R\bra{e} +\nn
                  &   g a\ket{0}_R\bra{e} + g^* a^\dagger\ket{e}\bra{0}_R. 
\end{align}
\color{black}
In the rotating wave approximation, in the basis 
$\{\ket{1_R,n-1},\ket{e, n-1 },\ket{0_R,n} \}$, the interaction Hamiltonian can be written as a 
direct sum,
\begin{align}
H_I(t) &= \bigoplus_n H^{(n)}(t), ~~
H^{(n)}(t)
=
\begin{pmatrix}
0 & \Omega(t)^* & 0 \\
\Omega(t) &  -\Delta & g\sqrt{n} \\
0 & g^*\sqrt{n} & 0
\end{pmatrix}.
\end{align}
Here $\Delta = \omega_L - \omega_0$ is the energy difference between the laser and the transition energy and itself can be a function of time.  
One of the eigenstates of $H^{(n)}(t)$ has a zero eigenvalue, $H^{(n)}(t)\ket{\psi_0(t)} = 0$, where
%
\mbox{$\ket{\psi_0(t)} = c\left(-r(t)\ket{1}_R\ket{n-1}+ 
\ket{0}_R\ket{n}\right)$}
, $r(t)   = \frac{g \sqrt{n}}{\Omega(t)}$,
and $c$ is a normalisation factor.

If $\Omega(t=0) \gg g$, then $\ket{\psi_0(t)} \approx \ket{0}_R$, i.e. $\ket{0}_R$ is the zero-eigenstate of $H_I(t)$. In our protocol, 
we initialise the atomic state as $\ket{0}_R$, then we adiabatically tune down $\Omega(t)$ such that at the end of the interaction ($t=T$), $\Omega(T) \ll g$. At this point, the zero-eigenstate $\ket{\psi_0(T)} \approx \ket{1}_R$.
\color{black}
That is, we have made a controlled spin population transfer from $\ket{0}_R$ to $\ket{1}_R$ depending on the presence of the photon. If the photon is absent, $\ket{0}_R\ket{\text{vac}}$ stays as $\ket{0}_R\ket{\text{vac}}$.

The joint state of the atom-photon evolves as $\rho(T)  = U_I(T) \rho(0)U_I^\dagger(T)$, 
$U_I(T) = \mathcal{T} \left\{ \int_0^T \exp(-i H_I t) dt \right\}$, where $\mathcal{T}\left\{\cdot \right\}$ is the time-ordering operator. Since we stay in the 0 eigenvalue of $H(t)$ for all time there is no dynamical phase accumulated. This is what makes STIRAP robust against timing errors.  
In the Appendix (including Refs.~\cite{PhysRevA.80.013417,shore2017picturing,PhysRevA.72.023405, RevModPhys.82.2313,timoney2011quantum,PhysRevLett.111.140501,koh2013high,you2011atomic,garcia2020overlapping,jessen2001quantum}), we discuss STIRAP in greater detail and show an explicit pulse that can complete the transfer while minimally populating $\ket{e}\bra{e}$.

\section{The encoder}
\label{sec:encoder}
Suppose we now prepare the register and the green ancilla (here the subscript $G$ denotes green) in the Bell state 
\begin{align}
\ket{\Phi_0}=\frac{1}{\sqrt2}(\ket{0_L}\ket{0_G} + \ket{1_L}\ket{1_G}).
\end{align}
Now, the red ancilla is initially prepared in state $\ket{0}_R$, so our set-up is in state
%
$\ket{\Psi_0} =\ket{\Phi_0} \otimes \ket{0}_R.$
%
Suppose Alice and Bob each have a copy of $\ket{\Psi_0}$, and they perform STIRAP individually (Fig.~\ref{f:memoryscheme} panel (iii)).
They share a single photon from the star
\mbox{$\frac{1}{\sqrt2}(\ket{1_p}_A\ket{\text{vac}}_B \pm e^{i \phi}\ket{\text{vac}}_A\ket{1_p}_B).$}
%
%
%
In the presence of the photon, the STIRAP interaction transforms $\ket{0}_R \rightarrow \ket{1}_R$, and the phase relationship in the photon is preserved.
This means that the state of the red ancillae (on Alice and Bob's sites) is now
\begin{align}
\frac{1}{\sqrt 2} \left(\ket{1_R, 0_R}_{AB} \pm e^{i \phi}\ket{0_R, 1_R}_{AB} \right).
\end{align}
Performing a Bell measurement on the red and green ancillae teleports the state onto the registers. After the Pauli operator correction dependent on the measurement outcome, the state of the registers between Alice and Bob becomes an entangled state, and the entanglement arises entirely from the starlight photon. 

Since the starlight state is mixed, after the encoding, the density matrix shared between Alice and Bob is
\begin{align} \label{eq:rhological}
\rho_{A B} \approx &(1-\epsilon)\ket{0_L 0_L}\bra{0_L 0_L}_{A B} + \epsilon \left(\frac{1+\gamma}{2}\right) \ket{\psi_{+,L}^\phi}\bra{\psi_{+,L}^\phi} \nn
& + 
\epsilon \left(\frac{1-\gamma}{2}\right)\ket{\psi_{-,L}^\phi}\bra{\psi_{-,L}^\phi} + O(\epsilon^2),
\end{align}
where $\ket{\psi^\phi _{\pm, L}} =(\ket{0_L, 1_L} \pm e^{i \phi}\ket{1_L, 0_L})/\sqrt 2.$

The states $\ket{\psi_{\pm,L}^\phi}$ are orthogonal to $\ket{0_L 0_L}$ and $\ket{1_L 1_L}$, and therefore can be distinguished via a parity measurement \cite{PhysRevLett.123.070504}. 
We can introduce additional pre-shared logical Bell pairs $\ket{\Phi^\pm} = (\ket{0_L,0_L} \pm \ket{1_L,1_L})/\sqrt2$,
which can be prepared by injecting a two-qubit Bell pair into Alice and Bob's QEC code by state injection \cite{ZLC00}.
The quality of the logical Bell pairs can be guaranteed by using distillation protocols \cite{campbell2017roads}.
Introducing additional pre-shared logical Bell pairs $\ket{\Phi^\pm} = (\ket{0_L,0_L} \pm \ket{1_L,1_L})/\sqrt2$, logical CZ gates between the memory qubits in $\rho_{AB}$ and $\ket{\Phi^+}$ can project out the vacuum:
\begin{align}
\rho_{A B} \otimes \ket{\Phi^+} &\stackrel{2 \times \text{CZ}}{\rightarrow} 
\ket{0_L,0_L}\bra{0_L,0_L}_{AB} \otimes\ket{\Phi^+}\bra{\Phi^+}+ \nn
&\epsilon \left(\frac{1+\gamma}{2}\right) \ket{\psi_{+,L}^\phi}\bra{\psi_{+,L}^\phi}\otimes \ket{\Phi^-}\bra{\Phi^-}  + \nn
 &\epsilon \left(\frac{1-\gamma}{2}\right)\ket{\psi_{-,L}^\phi}\bra{\psi_{-,L}^\phi} \otimes\ket{\Phi^-}\bra{\Phi^-}.
\end{align}

It suffices for Alice and Bob to perform local measurements to determine if the logical Bell pair is $\ket{\Phi^+}$ or $\ket{\Phi^-}$.
For instance, Alice and Bob would both
measure in the eigenbasis of the logical $X$ operator, and accept only the odd parity outcome.
This odd parity outcome corresponds to a projection onto the state $\ket{\Phi^-}$,
which reveals that a star photon has been captured.
This method extends to accommodate multiple photon events (Appendix).

After projecting out the vacuum, we can use local measurements to extract $\phi$ and $\gamma$. The QFI for $\phi$ is $\gamma^2 \epsilon$, and the QFI for $\gamma$ is $\epsilon/(1-\gamma^2)$.
 Local measurements are sufficient to saturate the quantum Cramer-Rao bound: indeed, projecting onto 
$\frac{1}{\sqrt2}(\ket{0_L}\pm e^{i \alpha}\ket{1_L})_A\otimes
\frac{1}{\sqrt2}(\ket{0_L}\pm \ket{1_L})_B$
is optimal, where $\alpha$ is an adjustable phase. Note that $\gamma$ and $\phi$ cannot be optimally simultaneously estimated (Appendix). 
If we want to avoid applying a logical phase gate, we could robustly do this using geometric phases during the STIRAP stage. By changing the relative phase of the pump pulse $\Omega(t)$ and the single atom coupling $g$ dynamically during the sequence, a geometric phase will accumulate depending on the path in parameter space \cite{PhysRevA.75.062302}.%

\section{Quantum error correction}
\label{sec:qec}
After STIRAP and the parity measurement, Alice and Bob share the quantum state $\rho'_{AB}$ in Eq.~\eqref{eq:rhological}, which is entangled over $2n$ qubits (they each hold $n$). We can calculate the QFI of $\rho'_{AB}$ with respect to the signal that has been encoded with QEC. 
For an [[$n,k,d$]] code, $n$ is the number of physical qubits, $k$ is the number of logical qubits encoded, and $d$ is the distance. The distance is the minimum number of physical errors it takes to change one logical codeword into another.
Any QEC code with distance $d$ can correct up to $t=\lfloor (d-1)/2 \rfloor$ errors \cite{roffe2019quantum}, $\lfloor \cdot \rfloor$ indicates the floor function.
 The choice of QEC depends on the number of available qubits and the noise model. We illustrate how our QEC scheme performs when a dephasing channel (see Appendix for depolarizing) afflicts each qubit. 
 If $n$ is small, the exact QFI can be calculated. We describe the dephasing channel as
\begin{align}
\mathcal{E}_
\text{dephase}{[\rho]} \rightarrow \left(1-p \right )\rho + p \sigma_z \rho \sigma_z^\dagger,
\end{align}
where $\sigma_z$ is the phase-flip operator. Here $p = 1/2$ corresponds to the completely dephasing channel. 

\begin{figure}[hbt]
\includegraphics[trim = 0cm 0.0cm 0cm 0cm, clip, width=1.0\linewidth]{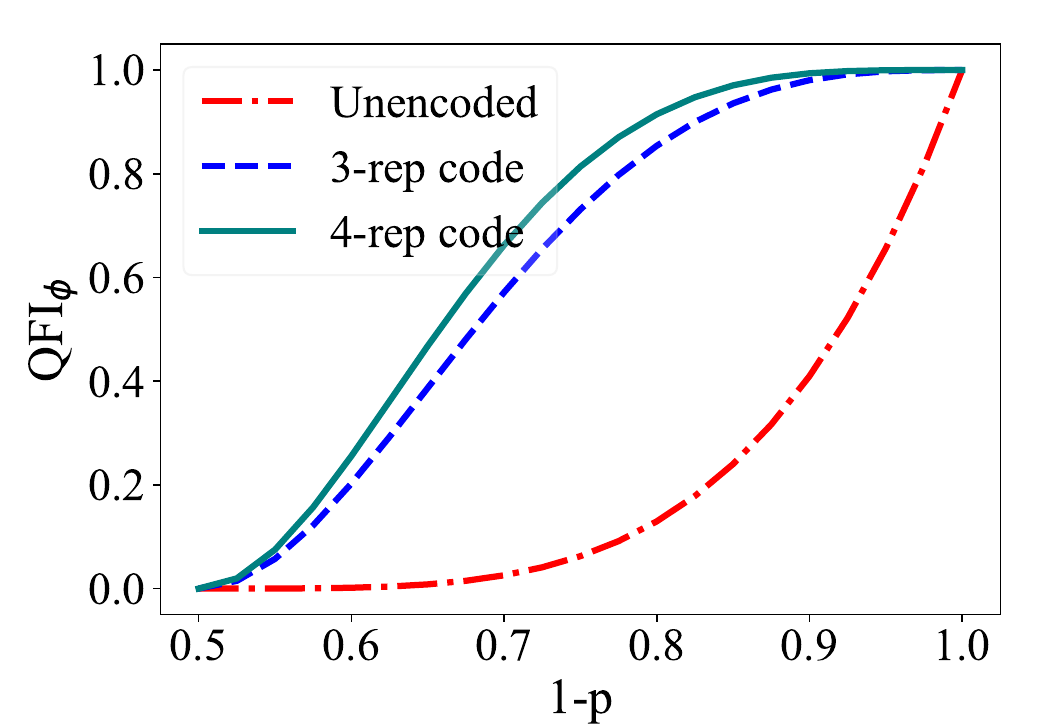} 
 \caption{\label{f:qec_phi} For $\gamma = 1$, the QFI for $\phi$ for the dephasing channel, when we use no encoding (red dotted-dashed line), the 3- and 4-repetition (blue dashed line and teal solid line).
 }
\end{figure}
In Fig.~\ref{f:qec_phi} we show the QFI of $\phi$ per photon as a function $p$. In most physical systems, dephasing is the dominant noise type. The unprotected case which does not use QEC codes has a QFI of $(1-2p)^4$, which drops off quickly even when $p$ is small. A simple quantum repetition code provides a significant advantage over the unprotected case for all values of dephasing. The logical states of the repetition codes are $\ket{0_L}=\ket{+}^{\otimes n}$ and $\ket{1}=\ket{-}^{\otimes n}$, $\ket{\pm} = (\ket{0} + \ket{1})/\sqrt 2$; and as $n$ increases, the state becomes more resilient.  In the limit of large $n$, the curve will approach a step function where the QFI is preserved up to $p<0.5$. This is evident from the Chernoff bound (below) because the phase-flip distance of the repetition code is $n$.

In Fig.~\ref{f:qec_gamma} we show the QFI of $\gamma$ per photon. 
Note $\gamma$ is a parameter associated with a non-unitary process, and behaves differently from $\phi$: 
the QFI can be preserved despite more than $(d-1)/2$ errors occurring. 
Surprisingly, if $\phi = 0$, the repetition code preserves its QFI perfectly. There are two cases leading to this phenomenon: if there are less than $n$ phase errors, the state is put onto a correctable subspace, and the corresponding normalised state has the same QFI as the original.
When $2n$ phase-flip errors occur, the logical states $(\ket{0_L} \ket{1_L} \pm \ket{1_L} \ket{0_L})/\sqrt2$ are eigenvectors of these errors with eigenvalues $\pm 1$: the state is invariant under the noise channel. 

\begin{figure}[hbt]
\includegraphics[trim = 0.0cm 0.0cm 0cm 0cm, clip, width=1.0\linewidth]{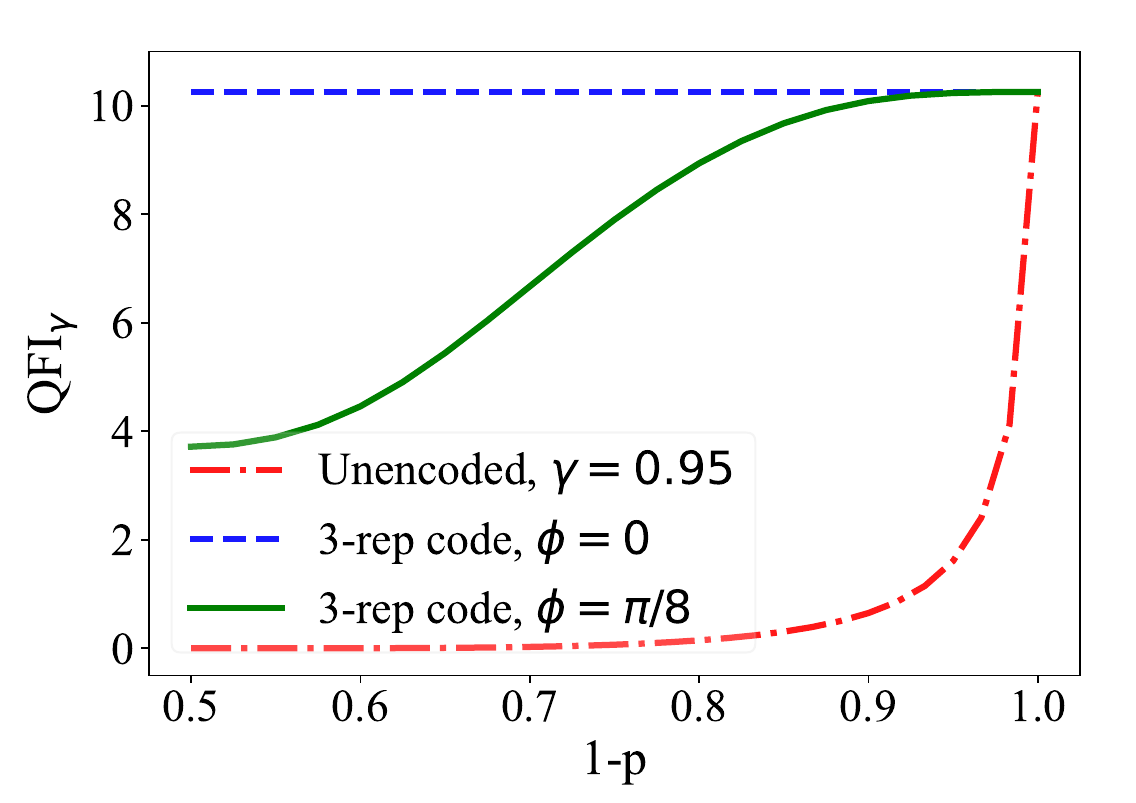} 
 \caption{\label{f:qec_gamma} For $\gamma = 0.95$,  the QFI for $\gamma$ of the dephasing channel, when no encoding is used (red dotted-dashed line), and the 3-repetition code, when $
 \phi = 0$ (blue dashed line), and when $\phi = \pi/8$ (green solid line).  }
\end{figure}

To understand the behaviour of large quantum codes, now let us consider any noise model that introduces error on each qubit independently with probability $p$.
Let $\epsilon_{\rm fail}$ denote the probability of having an uncorrectable error, where $\epsilon_{\rm fail}$ is at most the probability of having at least $d/2$ errors.
Using the Chernoff-Hoeffding bound for Bernoulli random variables \cite{Hoe63},
whenever $p<d/(2n)$, we have \cite{chernoff1952measure,okamoto1959some}
\begin{align}\label{eq:fail}
\epsilon_{\rm fail} 
\le 
e^{-D(p \| d/2n  ) n},
\end{align}
where $D(x\|y) = x \ln (x/y) + (1-x) \ln( (1-x)/(1-y)$ denotes the Kullback-Leibler divergence.
Note that $\epsilon_{\rm fail} $ vanishes exponentially in $n$ for small enough $p$. For large QEC codes, $d/(2n)$ asymptotes to a positive constant. By the quantum Gilbert-Varshamov bound \cite{FeM04,MaY08,JiX11,ouyang2014concatenated}, we know that if we use random QEC codes, we can have $d/n$ approaching $0.1893$ for large $n$. Hence, for our scheme, such QEC codes can tolerate noise afflicting up to $9.4\%$ of the qubits while preserving the QFI. See Appendix for further discussions, including Refs~\cite{LNCY97,ouyang2014permutation,PhysRevLett.120.050505,ALT01,CLX01,LXW09,PRXQuantum.2.030345,Cleland12,wu2022erasure,panteleev2022asymptotically,ouyang2021permutation,shibayama2021permutation,nakayama2020first,shibayama2021equivalence,baireuther2019neural,Chao2020.anyflag,PhysRevA.100.042332,campbell2017roads}).

\section{Discussions and conclusions}
We have proposed a general framework for applying QEC to an imaging task, where the experimenter did not prepare the probe.
Although we cannot illuminate our objects for astronomical imaging, 
we can nonetheless perform super-resolution imaging beyond the diffraction limit \cite{tsang2019resolving,PhysRevLett.117.190801,PhysRevLett.124.080503,PhysRevLett.127.130502}. 
From this perspective, our work complements the active research area of super-resolution imaging \cite{tsang2019resolving,PRXQuantum.2.030303,PhysRevLett.121.023904,PhysRevLett.121.180504,PhysRevLett.122.140505,PhysRevLett.118.070801,PhysRevLett.121.250503,hassett2018sub,zhou2019quantum}.

We have entered the stage where tens -- soon hundreds -- of qubits are becoming available.
 Much effort has focused on using noisy intermediate-scale quantum (NISQ) \cite{preskill2018quantum} devices to demonstrate capabilities that surpass classical computers. 
 We have proposed an application for a NISQ device for imaging. 
For the dominant noise type---dephasing---we show that a significant advantage can be gained even with a simple repetition code, and we can tolerate error rates up to 50$\%$. For noise types (even adversarial) that corrupt up to a certain fraction of the qubits we find the threshold of $9.4\%$ for which the QFI can be preserved. This threshold is significantly less stringent than that required for quantum computation.

\begin{acknowledgments}
ZH is supported by a Sydney Quantum Academy Postdoctoral Fellowship, and thanks Jonathan P. Dowling for inspiring this line of research. We thank Thomas Volz for his insightful discussions. G.K.B. acknowledges support from the Australian Research Council Centre of Excellence for Engineered
Quantum Systems (Grant No. CE 170100009).
Y.O. is supported by the Quantum Engineering Programme grant NRF2021-QEP2-01-P06, and also in part by NUS startup grants (R-263-000-E32-133 and R-263-000-E32-731), and the National Research Foundation, Prime Minister’s Office, Singapore and the Ministry of Education, Singapore under the Research Centres of Excellence program.
\end{acknowledgments}


%
\clearpage
\widetext
\appendix
\section{The model and the Quantum Cramer-Rao bound}

We model the incoming signal as a weak thermal state of light \cite{mandel1995optical} that has been multiplexed into frequency bands narrow enough for interferometry. 

For such an $n$-mode Gaussian state, for a particular band, given density matrix $ \rho$ and the creation/annihilation operators $ P =(a_1, a_1^\dagger,...,a_n, a_n ^\dagger)$, its properties are completely specified by the first and second moments
\begin{align}
\mu_k = \tr[P_k  \rho], \qquad \Sigma_{k,l} = \frac{1}{2}\tr[\{P_k - \mu_k, P_l - \mu_l\}\rho]
\end{align}
where $\{X,Y\} = XY +YX$ denotes the anticommutator.
 In basis $\{ a,  a^\dagger,  b, b^\dagger \} $, where $ a~ ( b)$ is the annihilation operator of the mode held by Alice (Bob), we have the first moments equal to 0, and the covariance matrix 
\begin{align}
\Sigma = \left(
\begin{array}{cccc}
 0 & \frac{\epsilon}{2}+\frac{1}{2} & 0 & \frac{1}{2} \gamma  \epsilon e^{i \phi } \\
 \frac{\epsilon}{2}+\frac{1}{2} & 0 & \frac{1}{2} \gamma  \epsilon e^{-i \phi } & 0 \\
 0 & \frac{1}{2} \gamma  \epsilon e^{-i \phi } & 0 & \frac{\epsilon}{2}+\frac{1}{2} \\
 \frac{1}{2} \gamma  \epsilon e^{i \phi } & 0 & \frac{\epsilon}{2}+\frac{1}{2} & 0 \\
\end{array}
\right),
\end{align}
here $\epsilon/2$ is the mean photon number of both modes.

The ultimate precision in the estimation is given by the quantum Cram\'er-Rao bound \cite{caves,caves1,giovannetti2011advances,giovannetti2006quantum}.
For the estimation of the parameter $\varphi$ encoded onto a quantum state $ \rho_\varphi$, this is a lower bound on the variance $(\Delta\hat\varphi)^2 = \langle \hat\varphi^2 \rangle - \langle \hat\varphi \rangle^2$ of any unbiased estimator $\hat\varphi$.
For unbiased estimators, the quantum Cramer-Rao bound establishes that 
\begin{align} \label{eq:var}
 (\Delta \hat \varphi) ^2 \geqslant \frac{1}{N} \frac{1}{J( \rho_\varphi)} \, ,
\end{align}
where $N$ is the number of probe systems used, and $J$ is the quantum Fisher information (QFI) associated with the global state $\rho_\lambda$ of the probes. 
The latter is defined as
\begin{eqnarray}
 J( \rho_\varphi)= \text{Tr}\left( L_\varphi^2   \rho_\varphi \right) \, ,
\end{eqnarray}
where $L_\varphi$ is the Symmetric Logarithmic Derivative (SLD) associated with the parameter $\varphi$ \cite{paris2009quantum}.
Consider a set of basis vectors $|e_1\rangle$, $|e_2\rangle$, $\dots$ in which $\rho_\varphi$ is diagonal:
\begin{align}
  \rho_\varphi = \sum_n p_n |e_n\rangle \langle e_n| \, .
\end{align}
The SLD is then given by
\begin{align}
 L_\varphi = 2 \sum_{n,m : p_n+p_m \neq 0} \frac{\braket{e_m|\partial_\varphi    \rho |e_n}}{p_n + p_m} \ket{e_m}\bra{e_n} \, ,
\end{align}
with $\partial_\varphi   \rho = \partial   \rho_\varphi/\partial \varphi$.
The quantum Cramer-Rao bound is asymptotically saturated in the limit that $N\rightarrow \infty$ \cite{PhysRevLett.110.240405}.

%
If multiple parameters are to be estimated, a necessary and sufficient condition for their joint optimal estimation is \cite{PhysRevA.94.052108}
\begin{align}
\tr(  \rho_{\boldsymbol{\varphi}}[L_{\varphi_i},L_{\varphi_j}] = 0).
\label{eq:joint}
\end{align}
The condition in Eq.~\eqref{eq:joint} is not satisfied for $\phi$ and $\gamma$, which means they will need to be separately estimated.

%
For conventional imaging, one typically characterises the resolution of an imaging system by the width of the point spread function (blurring on the image screen due to diffraction). The point spread function characterises how well one can localise a point source. For example, if we use a circular lens of diameter $D'$, and wavelength $\lambda$ then the point spread function if well-approximated by a Gaussian, whose standard deviation 
$\sigma = \frac{\sqrt 2 \lambda }{\pi ^2 D}$.

Its inverse
\begin{align}
\frac{1}{\sigma} = \frac{\pi  D'}{\sqrt{2} \lambda }
\end{align}
is proportional to the diameter and inversely proportional to the wavelength.

In the context of quantum imaging via parameter estimation, the metric for the resolution is the QFI, because it quantifies how well one can estimate the parameter of interest. In our case, by using a baseline $D$, the QFI for localising a point source is given by \cite{PhysRevLett.124.080503}

\begin{align}
QFI = \frac{4 \pi^2 D^2}{ \lambda^2}
\end{align}

The square root inverse of the QFI provides the best bound on the standard deviation of the estimate, in which case this case is $2\pi D/\lambda$. This is consistent with the usual notion of resolution.

\color{black}

\section{QFI of the stellar photon state }

In this section we derive the QFI associated with noise introduced by the environment in the unencoded case.  Consider the case where there is at most a single photon arriving on the two sites, where $\epsilon \ll1$. We can describe the optical state by the density matrix 
\begin{align}\label{eq:coupled}
\rho_{A B} \approx &(1-\epsilon)\ket{\text{vac},\text{vac}}\bra{\text{vac},\text{vac}}_{A_1 B_1} +  \epsilon \left(\frac{1+\gamma}{2}\right) \ket{\psi_+^\phi}\bra{\psi_+^\phi} + 
\epsilon \left(\frac{1-\gamma}{2}\right)\ket{\psi_-^\phi}\bra{\psi_-^\phi}
\end{align}
\noindent where $\ket{\psi_\pm^\phi} = 1/\sqrt{2}(\ket{1_p}_{A_1} \ket{\text{vac}}_{B_1} \pm e^{i\phi} \ket{\text{vac}}_{A_1} \ket{1_p}_{B_1})  $, here the subscript $p$ denotes a single photonic Fock state.

In the noiseless scenario, the QFI and SLD for $\gamma$ are:
\begin{align}
\text{QFI}_\gamma = &\frac{\epsilon }{1-\gamma ^2}, \nn
\text{SLD}_\gamma = &\frac{-1}{1-\gamma} \ket{\psi^\phi _-}\bra{\psi^\phi _-}+ 
                    \frac{1}{\gamma +1} \ket{\psi^\phi _+}\bra{\psi^\phi _+}.
\end{align}

And for $\phi$ 
\begin{align}
\text{QFI}_\phi &= \gamma^2 \epsilon, \nn
\text{SLD}_\phi &=  {i \gamma}(\ket{\psi^\phi _+}\bra{\psi^\phi _+}  
                           - \ket{\psi^\phi _-}\bra{\psi^\phi _-} ).
\end{align}

\subsection{Ideal case with QEC}
After being imprinted onto a QEC dcode, the density matrix shared between Alice and Bob is
\begin{align} \label{eq:rhological}
\rho_{A B} \approx &(1-\epsilon)\ket{0_L 0_L}\bra{0_L 0_L}_{A B} + \epsilon \left(\frac{1+\gamma}{2}\right) \ket{\psi_{+,L}^\phi}\bra{\psi_{+,L}^\phi}+
\epsilon \left(\frac{1-\gamma}{2}\right)\ket{\psi_{-,L}^\phi}\bra{\psi_{-,L}^\phi},
\end{align}
where $\ket{\psi^\phi _{\pm, L}} =(\ket{1_L, 0_L} \pm e^{i \phi}\ket{0_L, 1_L})/\sqrt 2.$

Once the vacuum is projected out, the state is
\begin{align}
\rho^{\prime}_{AB} = \left(\frac{1+\gamma}{2}\right) \ket{\psi_+^\phi}\bra{\psi_+^\phi} + 
 \left(\frac{1-\gamma}{2}\right)\ket{\psi_-^\phi}\bra{\psi_-^\phi}.
\end{align}

Local measurements are sufficient to saturate the quantum Cramer-Rao bound: indeed, projecting onto $1/\sqrt2 (\ket{0_L} \pm e^{i\alpha} \ket{1_L})_A \otimes 1\sqrt2 (\ket{0_L} \pm  \ket{1_L})_B $. The measurement outcomes are:
\begin{align} \label{eq:outcomes}
p(+,+) &= \frac{1}{4} (1+\gamma  \cos (\alpha -\phi )) \nn
p(+,-) &= \frac{1}{4} (1-\gamma  \cos (\alpha -\phi )) \nn
p(-,+) &= \frac{1}{4} (1-\gamma  \cos (\alpha -\phi )) \nn
p(-,-) &= \frac{1}{4} (1+\gamma  \cos (\alpha -\phi )).
\end{align}

The Fisher information for the parameter $\varphi$, for a set of measurement outcomes $i$ occurring with probability $p_i(\varphi)$ is given by
\begin{align}
I(\varphi) = \sum_i p_i(\varphi) \left(\frac{\partial \log p_i(\varphi)}{\partial \varphi}  \right)^2.
\end{align}

If we calculate the Fisher information associated with the measurement outcomes in Eq.~\eqref{eq:outcomes} for $\phi$, we have
\begin{align}
I(\phi) &= \frac{\gamma ^2 \sin ^2(\alpha -\phi )}{1-\gamma ^2 \cos ^2(\alpha -\phi )} \nn
        &= \gamma^2 \quad \text{     for } \quad  \alpha = \phi - \pi/2.
\end{align}
\color{black}
Now, for $\gamma$, we have
\begin{align}
I(\gamma) &= -\frac{\cos ^2(\alpha -\phi )}{\gamma ^2 \cos ^2(\alpha -\phi )-1} \nn
          &= \frac{1}{1-\gamma ^2}  \text{     for } \quad  \alpha = \phi.
\end{align}
The Fisher information for these parameters match the QFI, and hence these measurements saturate the quantum Cramer-Rao bound.

It is a common problem in quantum metrology that the optimal measurement, in general, can depend on the parameter itself. This is usually addressed in a feedback scenario, where the estimate of $\phi$ is updated as per measurement, and $\alpha$ is adjusted to our best estimate of $\phi$ \cite{PhysRevA.63.053804,PhysRevLett.85.5098,PhysRevA.95.053837}.

\color{black}

\subsection{Unprotected state with amplitude damping}

In the unencoded case, if the ground-excited state is used to store the information, the state of the memory qubit is predominantly subjected to dephasing and amplitude-damping noise.  We begin with the memory state
\begin{align}
\rho_{AB} = &(1-\epsilon) \ket{g,g}\bra{g,g}_{A_1 B_1} + \nn
&\epsilon\left(\frac{1+\gamma}{2}\right)\ket{\psi_+}\bra{\psi_+} +\nn
&\epsilon\left(\frac{1-\gamma}{2}\right)\ket{\psi_-}\bra{\psi_-}.
\end{align}

In the basis $\ket{g,e}$, the Amplitude damping channel (ADC) has Krauss operators 
\begin{align}
D_0 = \left(
\begin{array}{cc}
 1 & 0 \\
 0 & \sqrt{1-\eta } \\
\end{array}
\right), \quad
D_1 = \left(
\begin{array}{cc}
 0 & \sqrt{\eta } \\
 0 & 0 \\
\end{array}
\right).
\end{align}

After applying the ADC, the density matrix becomes
\begin{align}
\mathcal{E}{[\rho_{AB}]} \rightarrow 1-\epsilon&(1-\eta)\ket{g,g} + \nn
&(1-\eta)\epsilon\left(\frac{1+\gamma}{2}\right)\ket{\psi_+}\bra{\psi_+} +\nn
&(1-\eta)\epsilon\left(\frac{1-\gamma}{2}\right)\ket{\psi_-}\bra{\psi_-}.
\end{align}

That is, if the damping strength $\eta$ is equal on Alice and Bob's sites, then this is equivalent to reducing $\epsilon$ by a factor of $(1-\eta)$, and hence the QFI for both parameters is reduced by a factor $(1-\eta)$.

\section{Dephasing}

We now analyse the dephasing channel, where phase errors $\sigma_z$ occur with probability $p$. In our notation, $p=1/2$ corresponds to a completely dephasing channel that acts on a single-qubit state $\rho$
\begin{align}
\mathcal{E}{[\rho]} \rightarrow \left(1-{p} \right )\rho + {p} \sigma_z \rho  \sigma_z^\dagger.
\end{align}

Assuming the qubits held by Alice and Bob have the same dephasing parameter $p$, the QFI is
\begin{align}
\text{QFI}_\phi &= (1-2p)^4 \gamma^2 \epsilon \nn
\text{QFI}_\gamma &= \frac{(1-2p)^4 \epsilon }{1-\gamma ^2 (1-2p)^4}.
\end{align}

\section{FI and QFI bounds}

After QEC, we have the state
\begin{align}
\rho_{  \phi}=      (1-\epsilon_{\rm fail} ) \rho_\text{ideal}({  \phi})  + \epsilon_{\rm fail} \sigma,
\end{align}
where $\rho_\text{ideal}(  \phi) $ is the ideal state, and $\sigma$ is some noisy state.
Let $L$ be an observable that we care about.
Then we can calculate
\begin{align}
    \tr(\rho_{  \phi} L^2) - \tr(\rho_{  \phi} L)^2.
\end{align}
Using the error propagation formula,
the variance of the parameter we care about is 
\begin{align}
\Delta (\hat {  \phi})
    &= 
    \frac{\tr(\rho_{  \phi} L^2)- \tr(\rho_{  \phi} L)^2 }
    {\left|\frac{\partial}{\partial   \phi}   \tr(\rho_{  \phi} L) \right|^2}.\label{eq:error-propagation-1}
\end{align}

\section{Achievable bounds}

An explicit form of the optimal estimator for the parameter $\theta$ is given by \cite{paris2009quantum}
\begin{align}
\hat O_{  \phi} = {  \phi} \openone + \frac{\hat L_{  \phi}}{J(  \phi)}
\end{align}
where $L_\lambda$ is the SLD. This results in
\begin{align}
\braket{O_{  \phi}} = {  \phi},  \qquad
\braket{\Delta O}^2 = 1/J(  \phi).
\end{align}

For a particular observable $\hat X$, by error propagation, the achievable variance is

\begin{align}
(\Delta {  \phi})^2 = \frac{\tr [\rho_{  \phi} X^2] - \tr[\rho_{  \phi} X]^2}{|\frac{\partial}{\partial \phi} \tr[X\rho_{  \phi}]|^2}.
\label{eq:error-prop}
\end{align}

\subsection{Achievable bounds for an optimal observable based on the SLD - most reasonable noise channels}
Based on detection per photon (normalised by $1/\epsilon$), one optimal observable based on the SLD of $\phi$ is
\begin{align}
\hat P = (\ket{\psi^\alpha _+}\bra{\psi^\alpha _+}  
                           - \ket{\psi^\alpha _-}\bra{\psi^\alpha _-} ),
\end{align}
where $\ket{\psi_+^\alpha}=1/\sqrt{2}(\ket{1_L 0_L} + e^{i \alpha}\ket{0_L 1_L})$, and $\alpha$ is an adjustable phase.
Setting $\hat P =X$ in \eqref{eq:error-prop}, we find that the numerator of \eqref{eq:error-prop} is upper bounded by the norm of $P$, where $\|P\|^2 = 1$.

For the denominator, we start with
\begin{align}
 \rho_\phi P =& (1-\epsilon_\text{fail})  \rho_\text{ideal}(\phi) P  + \epsilon_\text{fail} \sigma P \nn
\braket{ \rho_\phi P}   =& (1-\epsilon_\text{fail}) \gamma \cos(\alpha +\phi) + \braket{\epsilon_\text{fail} \sigma P}
\end{align}
\noindent where the first term stems from the fact that, if the error is correctable, the state can be returned to being $\rho_\text{ideal}$, and
$\rho = (\frac{1 + \gamma}{2})\ket{\psi_+^\phi}\bra{\psi_+^\phi} + (\frac{1 - \gamma}{2})\ket{\psi_-^\phi}\bra{\psi_-^\phi}$.

From the reverse triangle inequality, $|x-y| \geq  \big| |x| - |y| \big|$, we have
\begin{align}
|\partial_\phi\braket{ \rho_\phi P}| &\geq  \gamma (1-\epsilon_\text{fail})|  \sin(\alpha +\phi)| - \epsilon_\text{fail} |\partial_\phi \braket{ \sigma P}| .
\end{align}

The magnitude of $\braket{ \sigma P} \leq \gamma$, because $P$ is an operator that picks out the real component of the off-diagonal components in the computational basis (logical). For most ``reasonable" noise channels (i.e. most of the physical noise types considered in the QEC literature), the dependence of $|\braket{\sigma P}|$ on $\phi$ should not be larger than that of $\braket{\rho_\text{ideal}(\phi) P}$, which means that the derivative has magnitude at most equal to $ \sin(\beta + \phi)]$, where $\beta$ can be in general not equal to $\alpha$. Assuming $\epsilon_\text{fail} \leq 1/2$,  when we adjust the phase in the measurement $\alpha$ such that
$\alpha -\phi = \pi/2$, we have
\begin{align}
|\braket{ \rho_\phi P}| \geq \gamma (1-2\epsilon_\text{fail})^2,
\end{align}
and therefore
\begin{align}
(\Delta \phi)^2 \leq \frac{1}{\gamma^2 (1-2\epsilon_\text{fail})^2}.
\end{align}

In fact, for local channels such as depolarising, dephasing and amplitude damping, any noisy component of the state has no off-diagonal components in the computational basis, where
$\braket{\sigma P} = 0$. In these cases, we achieve
$(\Delta \phi)^2 \leq {1}/{\gamma^2 (1-\epsilon_\text{fail})^2}.$

\subsection{Achievable bounds for an observable based on local measurements - most reasonable noise channels}
Now, consider the separable observable, where the state is projected onto the basis $\ket{\pm^\alpha}_A\otimes \ket{\pm}_B =\frac{1}{\sqrt2}(\ket{0} \pm e^{i \alpha }\ket{1})_A\otimes\frac{1}{\sqrt2}(\ket{0} \pm \ket{1})_B$,
and consider the observable
\begin{align}
P_\text{sep} = \ket{+^\alpha,+}\bra{+^\alpha,+} +\ket{-^\alpha,+}\bra{-^\alpha,+} - ( \ket{+^\alpha,-}\bra{+^\alpha,-} +\ket{-^\alpha,+}\bra{-^\alpha,+}).
\end{align}

We have $\|P_\text{sep}\|^2 = 1$.
Note that 
\begin{align}
\rho_\phi P_\text{sep} =& (1-\epsilon_\text{fail})  \rho_\text{ideal}(\phi) P_\text{sep}  + \epsilon_\text{fail} \sigma P_\text{sep} \nn
\braket{\rho_\phi P_\text{sep}}   = & (1-\epsilon_\text{fail}) \gamma  \cos (\alpha +\phi ) + \braket{\epsilon_\text{fail} \sigma P_\text{sep}}.
\end{align}
Similarly, as before, we obtain an upper bound for $(\Delta \phi)^2$.

\section{STIRAP - a brief review and pulse optimization}

\begin{figure}[h!]
\includegraphics[trim = 0cm 0.0cm 0cm 0cm, clip, width=0.3\linewidth]{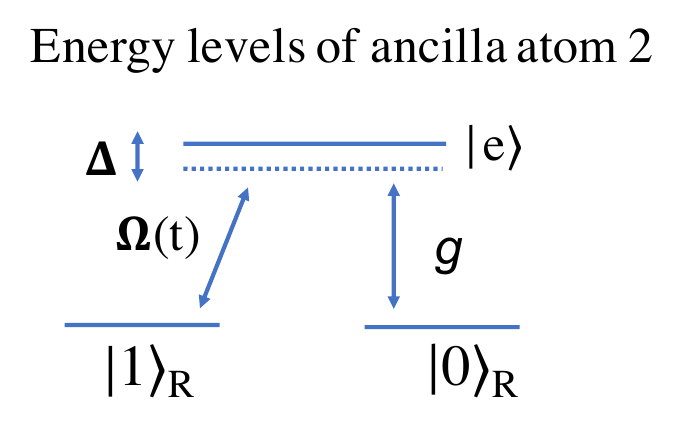} 
 \caption{\label{f:stirap_levels_supp} The linkage pattern showing the fields $\Omega(t)$ linking $\ket{1}_R$ to $\ket{e}$, the field $g$ linking $\ket{0}_R$ to $\ket{e}$, and the single-photon detuning $\Delta$. }
\end{figure}

STIRAP originated as a technique for transferring population between two quantum states by coupling them with two fields via an intermediate state, the example in the main text is given in Fig.~\ref{f:stirap_levels_supp}.

STIRAP allows, in principle, the complete transfer of population along a three-state chain $\ket{0}_R\rightarrow\ket{e}\rightarrow\ket{1}_R$ from an initially populated quantum state $\ket{0_R}$ to a target quantum state $\ket{1}_R$, induced by a light-matter interaction Hamiltonian fields that couples the intermediate $\ket{e}$ to states $\ket{0}_R$ and $\ket{1}_R$.

In this paper, we define $\Omega(t)$ to be the coupling strength between $\ket{1}_R$ and $\ket{e}$, and corresponds to
the strength of the pump laser. The coupling between $\ket{0}_R$ and $\ket{e}$ is denoted $g$, which is the atom-cavity coupling.

Defining $n$ to be the number of photons in the cavity, the STIRAP Hamiltonian is
\begin{align}
H_\text{stirap}(t) = &~ \omega_0 \ket{e}\bra{e} + \omega_c a^\dagger a  + \nn
                 & \Omega(t) e^{-i \omega_L t} \ket{e}\bra{1} + \Omega(t)^*e^{i \omega_L t}\ket{1}\bra{e} +\nn
                  &   g a\ket{0}\bra{e} + g^* a^\dagger\ket{e}\bra{0}. 
\end{align}
In the rotating wave approximation, in the basis 
$\{\ket{1_R,n-1},\ket{e, n-1 },\ket{0_R,n} \}$, the interaction Hamiltonian can be written as a 
direct sum,
\begin{align} \label{eq:stirap_H}
H_I(t) &= \sum_n H^{(n)}(t), ~~
H^{(n)}(t)
=
\begin{pmatrix}
0 & \Omega(t)^* & 0 \\
\Omega(t) &  -\Delta & g\sqrt{n} \\
0 & g^*\sqrt{n} & 0
\end{pmatrix}.
\end{align}
Here $\Delta = \omega_L - \omega_0$ is the energy difference between the laser and the transition energy and itself can be a function of time.



The time-varying fields $\Omega(t)$ and $g$ describe the couplings between the intermediate state $\ket{e}$ and respectively, the initial state $\ket{1}_R $ and $\ket{0}_R$. STIRAP can be explained via the so-called ``dark state", which is a zero-eigenvalue eigenstate of Eq.~\eqref{eq:stirap_H}: $H^{(n)}(t)\ket{\psi_0(t)} = 0$, where

\begin{align}\label{eq:zeroeigen}
\ket{\psi_0(t)} &= c\left(-r(t)\ket{1}_R\ket{n-1}+ \ket{0}_R\ket{n}\right), \nn
 r(t)   &= \frac{g \sqrt{n}}{\Omega(t)}
\end{align}
and $c$ is a normalisation factor.

 The pulses in STIRAP act in a seemingly counterintuitive ordering. The field $\Omega(t)$, though acting on the unpopulated states $\ket{e}$ and $\ket{1}_R$, must act before (but overlapping) with the field $g$ \cite{RevModPhys.89.015006,PhysRevA.80.013417}. In our setting, $g$ is the coupling between the photon and the cavity and is therefore constant. Thus $\Omega(t)$ must be much stronger than $g$ before the interaction. I.e., if we have

 \begin{align} 
\lim_{t=0} \frac{|g|}{|\Omega(t)|} \approx 0, \qquad  \lim_{t=T} \frac{|\Omega(t)|}{|g|} \approx 0
\end{align}
then before the interaction, $\ket{\psi_0(t=0)} \approx \ket{0}_R\ket{n=1}$, and at the end of the interaction $\ket{\psi(T)} \approx \ket{1}_R\ket{n=0}$.

That is, in the presence of a photon, $\ket{0}_R$ is coupled to $\ket{e}$, and if we adiabatically tune down $\Omega(t)$ such that at the end of the interaction, $t=T$, we have made a controlled spin population transfer from $\ket{0}_R$ to $\ket{1}_R$ depending on the presence of the photon.
If the evolution is indeed adiabatic, then the population passes from state $\ket{0}_R$ to state $\ket{1}_R$. Moreover, because the dark state does not contain a contribution from the excited intermediate state $\ket{e}$, we can avoid spontaneous emission losses.

On the other hand, if no photon is present, then the coupling between $\ket{0}_R$ and $\ket{e}$ is zero, and
 $\ket{0}_R\ket{\text{vac}}$ stays as $\ket{0}_R\ket{\text{vac}}$.

In comparison to STIRAP used for other applications, our control parameters are more restrained - "normally",  both the fields $\Omega$ and $g$ would be applied using a laser, and ideally both the pulses are Gaussian \cite{shore2017picturing}. Here the coupling $g$ is constrained to be constant.

\begin{figure}[h!]
\includegraphics[trim = 0cm 0.0cm 0cm 0cm, clip, width=0.5\linewidth]{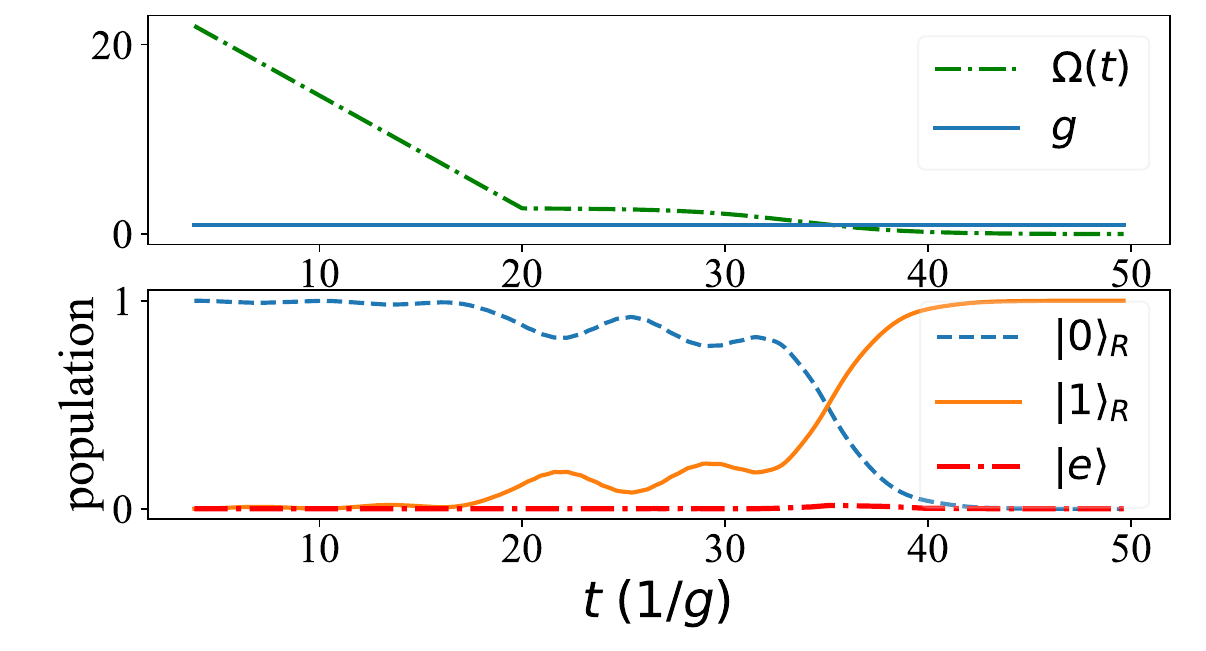} 
 \caption{\label{f:population_transfer} \textbf{Population transfer for of a three-state adiabatic passage}. Top: the interaction strengths of $g$ and $\Omega$ as a function of time $t$ in units of $1/g$. Bottom: occupancy of in $\ket{e},\ket{0}_R$ and $\ket{1}_R$ in ancillary atom 2 ($R$). The detuning parameter is set to $\Delta(t) = g^2 +\Omega^2(t) $ to satisfy the adiabatic condition \cite{PhysRevA.80.013417}. Note that the excited state $\ket{e}\bra{e}$ is minimally populated.}
\end{figure}

\color{black}
In Fig.~\ref{f:population_transfer}, for $n=1$, we show that the population transfer between the atomic states $\ket{0}_R$ and $\ket{1}_R$ can be completed while {minimally} populating the excited state $\ket{e}$, thus avoiding spontaneous emission. We have fixed the parameter $g=1$, and $\Omega(t)$ is numerically optimised to maximise the transfer from $\ket{0}_R$ to $\ket{1}_R$. The pulse is divided into three intervals for $\Omega(t)$: during the first interval, $\Omega(t)$ is linear, in the second interval, it is a hyperbolic tangent function, and finally, a short linear taper to ensure $\Omega(T)=0$.
During this STIRAP pulse sequence, the maximum of the density matrix element $\ket{e}\bra{e}$ is $1.6\%$ around $T\approx 35$.

Physically, STIRAP has been implemented on many different platforms, including Rydberg atoms\cite{PhysRevA.72.023405, RevModPhys.82.2313}, trapped ions \cite{timoney2011quantum,PhysRevLett.111.140501} semiconductor quantum dots \cite{koh2013high}, superconducting circuits \cite{you2011atomic} and etc. The metric for the quality of the atom-cavity system
is the cooperativity $C = 2 g^2/\kappa \eta $, which is the ratio of $g$ to the cavity loss rate $\kappa$, and the decay rate of the atom into non-cavity modes $\eta$. 
The potential for achieving high cooperativity gives cavity QED a central role in the development of high-fidelity quantum gates. A recent result reports a cooperativity $C=299$ \cite{garcia2020overlapping}
 for Rubidium atoms in a fibre cavity that is potentially compatible with our setup.

 During the STIRAP pulse, there is a possibility of dephasing on the red ancilla qubits. This dephasing can be mitigated by a judicious choice of ground states. For example, in an alkali atom with hyperfine structure and half-integer nuclear spin $(I>1)$ one could choose the states $\ket{0}_R=\ket{F_{\uparrow},M_F=1}$, $\ket{1}_R=\ket{F_{\downarrow},M_F=-1}$, and $\ket{e}=\ket{F',M_F=0}$. Because the Land\'e g-factors are equal and opposite for the two ground-state hyperfine manifolds, the shift due to extraneous magnetic fields will be zero to first order. Additionally, with this choice of levels, a time-dependent detuning $\Delta(t)$ to satisfy adiabaticity can be realized by turning on a time-dependent magnetic field along the quantization axis of the atom \cite{jessen2001quantum}.
\color{black}

\section{Higher photon contributions}

In reality, the state received from the star is thermal. Therefore the probability of having more than 1 photon is non-zero. Since the mean photon number is usually low, we truncate the term at two photons; 
in this section, we work out the two-photon states in the Fock basis.

The covariance matrix in the basis $ \{a, a^\dagger, b, b^\dagger \} $ ($a ~(b)$ being the mode at the collector held by Alice (Bob) is 
\begin{align}
\Sigma = \left(
\begin{array}{cccc}
 0 & \frac{\epsilon}{2}+\frac{1}{2} & 0 & \frac{1}{2} \gamma  \epsilon \exp (i \phi ) \\
 \frac{\epsilon}{2}+\frac{1}{2} & 0 & \frac{1}{2} \gamma  n \exp (-i \phi ) & 0 \\
 0 & \frac{1}{2} \gamma  \epsilon \exp (-i \phi ) & 0 & \frac{n\epsilon}{2}+\frac{1}{2} \\
 \frac{1}{2} \gamma  \epsilon \exp (i \phi ) & 0 & \frac{\epsilon}{2}+\frac{1}{2} & 0 \\
\end{array}
\right)
\end{align}
given $\epsilon$ is the mean total photon number, and $\gamma,\phi$ as per defined above.

This state is diagonalisable with a phase shifter and a 50:50 BS:
\begin{align}
S =
\frac{1}{\sqrt2}\left(
\begin{array}{cccc}
 1 & 0 & 1 & 0 \\
 0 & 1 & 0 & 1 \\
 1 & 0 & -1 & 0 \\
 0 & 1 & 0 & -1 \\
\end{array}
\right) \left(
\begin{array}{cccc}
 1 & 0 & 0 & 0 \\
 0 & 1 & 0 & 0 \\
 0 & 0 & \exp (-i \alpha ) & 0 \\
 0 & 0 & 0 & \exp (i \alpha ) \\
\end{array}
\right).
\end{align}

We have
\begin{align}
\Sigma' =
S \Sigma S^T = \left(
\begin{array}{cccc}
 0 & \frac{1}{2} (\gamma  \epsilon+\epsilon+1) & 0 & 0 \\
 \frac{1}{2} (\gamma  \epsilon+\epsilon+1) & 0 & 0 & 0 \\
 0 & 0 & 0 & \frac{1}{2} (-\gamma  \epsilon+\epsilon+1) \\
 0 & 0 & \frac{1}{2} (-\gamma  \epsilon+\epsilon+1) & 0 \\
\end{array}
\right).
\end{align}

We will need the Hermitian conjugate of $S$
\begin{align}
S^{\dagger} = \frac{1}{\sqrt 2}
\left(
\begin{array}{cccc}
 1 & 0 & 1 & 0 \\
 0 & 1 & 0 & 1 \\
 e^{i \alpha } & 0 & -e^{i \alpha } & 0 \\
 0 & e^{-i \alpha } & 0 & -e^{-i \alpha } \\
\end{array}
\right)
\end{align}
because a thermal state in the number basis is written as
\begin{align}
\rho_G = S^\dagger \left( \bigotimes^n   \rho_n \right)S.
\end{align}

This means that for a two-mode thermal state, we have
\begin{align}\label{eq:twomode}
\rho &= \rho_a \otimes \rho_b, \nn
 \rho_a &= \sum_{i}\frac{1}{n_a +1} \left(\frac{n_a}{ n_a +1}\right)^{i}\ket{i}\bra{i}, 
            \qquad n_a = \frac{1}{2}(\epsilon + \gamma \epsilon)  \nn
 \rho_b &= \sum_{j}\frac{1}{n_b +1} \left(\frac{n_b}{ n_b +1}\right)^{j}\ket{j}\bra{j}, 
            \qquad n_b = \frac{1}{2}(\epsilon - \gamma \epsilon)  \nn
\rho_a \otimes \rho_b &= \sum_{i,j} \frac{1}{(n_a + 1)(n_b +1)} 
          \left(\frac{n_a}{n_a +1}\right)^{i} 
          \left(\frac{n_b}{n_b +1}\right)^{j} 
          \frac{ ({a'}_1^{\dagger})^i}{i!} 
          \frac{ ({a'}_2^{\dagger})^j}{j!}\ket{0,0}\bra{0,0}(a'_1)^i (a'_2)^j.
\end{align}
To invert the diagonalising operation, the operators in Eq~\eqref{eq:twomode} should transform as
\begin{align}
{a'}^\dagger_1 &\rightarrow \frac{1}{\sqrt2}(a_1^\dagger + a_2^\dagger e^{i \phi}) \nn
{a'}^\dagger_2 &\rightarrow \frac{1}{\sqrt2}(a_1^\dagger  - a_2^\dagger e^{i\phi}).
\end{align}

Now denote $p(i,j)$ to have $i$ and $j$ number of photons from the modes held by Alice and Bob after diagonalising the covariance matrix.
Let us separate the terms. For a total of 0 photons, we have

\begin{align}
p(0,0) = \frac{1}{(n_a +1)(n_b +1)}.
\end{align}

With one photon shared between either mode, we have
\begin{align} \label{eq:rho1}
p(1) =& p(1,0) + p(0,1).\nn
\rho^{(1)} = &\frac{1}{(n_a +1)(n_b +1)} \left(\frac{n_a}{n_a+1} \right)
     \frac{ (a_1^\dagger + e^{i\phi}a_2^\dagger)}{2 } \ket{0,0}\bra{0,0}(a_1 + e^{-i\phi}a_2) +  \nn
       & \frac{1}{(n_a + 1)(n_b +1)} 
          \left(\frac{n_b}{n_b +1}\right)
          \frac{ ( a_1^\dagger  - a^\dagger_2 e^{i \phi})}{2}\ket{0,0}\bra{0,0}( a_1  - a_2 e^{i \phi}) \nn
&= \frac{1}{(n_a+1 )(n_b+1)}\frac{1}{2}\left( \frac{n_a}{n_a+1} \right)\ket{\psi_+^\phi}\bra{\psi_+^\phi}
+\frac{1}{(n_a+1 )(n_b+1)}\frac{1}{2}\left( \frac{n_b}{n_b+1} \right)\ket{\psi_-^\phi}\bra{\psi_-^\phi}.
\end{align}

Now, for the two-photon case, we have $i=1, j=1$, $i = 2, j =0$, or $i = 0, j =2$.

\begin{align}
p(2) = &p(1,1) + p(2,0) + p(0,2)\nn
 \rho^{(2)}    =&\frac{1}{(n_a + 1)(n_b +1)} 
          \left(\frac{n_a}{n_a +1}\right)
          \left(\frac{n_b}{n_b +1}\right)
           ({a'}_1^{\dagger})
           ({a'}_2^{\dagger})\ket{0,0}\bra{0,0}(a'_1) (a'_2) + \nn
  &\frac{1}{(n_a + 1)(n_b +1)} \left(\frac{n_a}{n_a +1}\right)^2
  \frac{({a'}_1^\dagger )^2}{2!} \ket{0,0}\bra{0,0}(a'_1)^2     + 
  \frac{1}{(n_a + 1)(n_b +1)} \left(\frac{n_b}{n_b +1}\right)^2
  \frac{({a'}_2^\dagger )^2}{2!} \ket{0,0}\bra{0,0}(a'_2)^2 .
\end{align}

We evaluate the following expression,
\begin{align}
&({a'}_1^{\dagger})({a'}_2^{\dagger})\ket{0,0}\bra{0,0}(a'_1) (a'_2) \nn
=& \frac{1}{4}(a^\dagger_1 + a^\dagger_2 e^{i \phi})
              (a^\dagger_1 - a^\dagger_2 e^{i \phi}) \ket{0,0}\bra{1,1}
              (a_1 + a_2 e^{-i \phi})
              (a_1 - a_2 e^{-i \phi}) \nn 
=& \frac{1}{4}(a_1^{\dagger 2} -e^{i 2\phi} a_2^{\dagger2}  )   \ket{0,0}\bra{0,0}
                (a_1^{ 2} -e^{-i 2\phi} a_2^{2}  )\nn
=&\frac{(\ket{2,0}-e^{i 2\phi}\ket{0,2})}{\sqrt2}\frac{(\bra{2,0}-e^{-i 2\phi}\bra{0,2})}{\sqrt 2} \nn
=&\ket{\Psi^2_0}\bra{\Psi^2_0}.
\end{align}

For the $i = 2, j=0$ term,
\begin{align}
 &\frac{({a'}_1^\dagger )^2}{2!} \ket{0,0}\bra{0,0}(a'_1)^2   \nn
=&\frac{1}{2!} \frac{1}{2} (a_1^\dagger + a_2 ^\dagger e^{i\phi})^2 \ket{0,0}\bra{0,0}
               \frac{1}{2} (a_1 + a_2 e^{-i\phi})^2 \nn
=& \frac{1}{16}(a_1^{\dagger2} + 2 a_1^\dagger a_2^\dagger e^{i\phi} + a_2^{\dagger 2}e^{i 2\phi})\ket{0,0}\bra{0,0}(h.c) \nn
=&\frac{1}{16}( \sqrt{2}\ket{2,0} + 2e^{i\phi}{1,1}  + \sqrt{2}e^{i2\phi}\ket{0,2} )\otimes h.c \nn
=& \frac{1}{2}  \times\frac{(\ket{2,0}+\sqrt{2}e^{i\phi}\ket{1,1} +e^{i 2\phi}\ket{0,2})}{2}\otimes h.c \nn
=& \frac{1}{2}\ket{\Psi^2_+}\bra{\Psi^2_+}.
\end{align}
Likewise for the $i=0, j = 2$ term, we will have
\begin{align}
&\frac{({a'}_2^\dagger )^2}{2!} \ket{0,0}\bra{0,0}(a'_2)^2   \nn
&=\frac{1}{2}  \times\frac{(\ket{2,0}-\sqrt{2}e^{i\phi}\ket{1,1} +e^{i 2\phi}\ket{0,2})}{2}\otimes h.c. \nn
=& \frac{1}{2}\ket{\Psi^2_-}\bra{\Psi^2_-}.
\end{align}

Overall, the two-photon contributions are
\begin{align}\label{eq:rho_two_photons}
 \rho^{(2)} =&\frac{1}{(n_a + 1)(n_b +1)} 
          \left(\frac{n_a}{n_a +1}\right)
          \left(\frac{n_b}{n_b +1}\right) 
          \ket{\Psi^2_0}\bra{\Psi^2_0} + \nn
&\frac{1}{(n_a + 1)(n_b +1)} \left(\frac{n_a}{n_a +1}\right)^2\frac{1}{2} \ket{\Psi^2_+}\bra{\Psi^2_+} + 
 \frac{1}{(n_a + 1)(n_b +1)} \left(\frac{n_b}{n_b +1}\right)^2\frac{1}{2} \ket{\Psi^2_-}\bra{\Psi^2_-}.
\end{align}

\section{Modified Hamiltonian due to multiple photons}

Given $n$ photons in the cavity, the full STIRAP Hamiltonian is
\begin{align}
H_\text{stirap} = &~ \omega_0 \ket{e}\bra{e} + \omega_c a^\dagger a  + \nn
                  &\Omega e^{-i \omega_L t} \ket{e}\bra{1} + \Omega^*e^{i \omega_L t}\ket{1}\bra{e} +\nn
                  &   g a\ket{0}\bra{e} + g^* a^\dagger\ket{e}\bra{0}  .
\end{align}

In the rotating wave approximation
$e^{-i\theta a^\dagger a} a e^{i\theta a^\dagger a} = e^{i\theta} a$, 
\begin{align}
 H_I  &= U^\dagger_c H U_c -H_c, \nn
    H_c &= \omega_c a^\dagger a + \omega_L \ket{e}\bra{e},\nn
    U_c &= e^{-i\omega_c a^\dagger a t} e^{-i \omega_L t \ket{e}\bra{e}}  \nn
 H_I = &-\Delta \ket{e}\bra{e}  + 
                   \Omega\ket{e}\bra{1} +
                   \Omega^* \ket{1}\bra{e} +
                   g a  \ket{0}\bra{e} + 
                   g^* a^\dagger  \ket{e}\bra{0},
\end{align}
where $\Delta = \omega_L -\omega_0$.
The last two terms connect pairs of bases with the same excitation number, i.e. $\ket{g,n} \rightleftharpoons\ket{e,n-1}$.
This means that the Hamiltonian can be written as a direct sum in the basis
\begin{align} 
\{\ket{1,n-1},\ket{e, n-1 },\ket{0,n} \}.
\end{align}

We have ($H_{i,j} = \braket{i|H|j}$)
\begin{align}
H^{(n)}&=
\begin{pmatrix}
\braket{1,n-1|H_\text{stirap}|1,n-1} & \braket{1,n-1|H_\text{stirap}|e,n-1} & \braket{1,n-1|H_\text{stirap}|0,n}   \nn
\braket{e,n-1|H_\text{stirap}|1,n-1} & \braket{e,n-1|H_\text{stirap}|e,n-1} & \braket{e,n-1|H_\text{stirap}|0,n}  \nn
\braket{0,n|H_\text{stirap}|1,n-1} & \braket{0,n|H_\text{stirap}|e,n-1} & \braket{0,n|H_\text{stirap}|0,n}
\end{pmatrix} \nn
&=
\begin{pmatrix}
0 & \Omega^* & 0 \nn
\Omega &  -\Delta & g\sqrt{n} \nn
0 & g^*\sqrt{n} & 0
\end{pmatrix}.
\end{align}

If the transfer unitary is perfect, the initial state turns into (with some phase difference, which we will observe numerically)
\begin{align}
\ket{0,0}_A\ket{0,2}_B &\rightarrow \ket{0,0}_A\ket{1,1}_B \nn
%
\ket{0,1}_A\ket{0,1}_B &\rightarrow \ket{1,0}_A\ket{1,0}_B.
\end{align}

Therefore if our initial state is
\begin{align}
\ket{0}_A\ket{0}_B \otimes 
   &\frac{1}{2}(\ket{2,0} + \sqrt{2} e^{i \phi}\ket{1,1} + e^{i 2\phi}\ket{0,2}) \nn
= &\frac{1}{2}\ket{0,2}_A \ket{0,0}_B + 
  \frac{1}{\sqrt2}e^{i\phi}\ket{0,1}_A\ket{0,1}_B + 
  \frac{1}{2}e^{i2\phi}\ket{0,0}_A \ket{0,2}_b 
  \end{align}
 then STIRAP takes it to the state
  \begin{align}
\rightarrow &
\frac{1}{2}\ket{1,1}_A \ket{0,0}_B + 
  \frac{1}{\sqrt2}e^{i(\phi + \delta)}\ket{1,0}_A\ket{1,0}_B + 
  \frac{1}{2}e^{i2\phi}\ket{0,0}_A \ket{1,1}_B
\end{align}
We observe an additional relative phase on the $\ket{1,0}\ket{1,0}$ term, due to the dispersive coupling of the atom to the cavity in the presence of an additional photon.

Now, we are faced with the task of distinguishing between the different cases without perturbing the encoded phase information. Using the subscripts $l$ to denote levels of an atomic system and $p$ for photonic Fock states. These are the three states:

a. No photon has arrived:
\begin{align}
\ket{0_l,0_p}_A\ket{0_l,0_p}_B.
\end{align}

b. A single photon has arrived:
\begin{align}
\frac{1}{\sqrt2}( \ket{1_l,0_p}_A \ket{0_l,0_p}_B 
   \pm e^{i\phi}  \ket{0_l,0_p}_A \ket{1_l,0_p}_B ).
\end{align}

c. Two photons have arrived:
\begin{align}
\ket{\psi_1} = \frac{1}{\sqrt 2}&(\ket{1_l,1_p}_A\ket{0_l,0_p}_B-e^{i 2\phi}\ket{0_l,0_p}_A\ket{1_l,1_p}_B) \nn
\ket{\psi_2} \rightarrow \frac{1}{2}&\ket{1_l,1_p}_A \ket{0_l,0_p}_B + 
  \frac{1}{\sqrt2}e^{i(\phi + \delta)}\ket{1_l,0_p}_A\ket{1_l,0_p}_B + 
  \frac{1}{2}e^{i2\phi}\ket{0_l,0_p}_A \ket{1_l,1_p}_B\nn
\ket{\psi_3 }\rightarrow  \frac{1}{2}&\ket{1_l,1_p}_A \ket{0_l,0_p}_B - 
  \frac{1}{\sqrt2}e^{i(\phi + \delta)}\ket{1_l,0_p}_A\ket{1_l,0_p}_B + 
  \frac{1}{2}e^{i2\phi}\ket{0_l,0_p}_A \ket{1_l,1_p}_B.
\end{align}

If we perform parity checks on the memory qubits and the cavity photon states (which can be achieved by using another auxiliary atom). We assume the Bell pairs 
$\ket{\Phi^+}$ are shared between Alice and Bob, where the subscript denotes which system they are paired with. For example,
 \begin{align}
\ket{1_l}_A\ket{0_l}_B \ket{\Phi^+_l}\stackrel{2 \times \text{CZ}}{\rightarrow} \ket{1_l}_A\ket{0_l}_B \ket{\Phi^-_l}
\end{align}
 denotes that 2 CZ gates (one on Alice's and one on Bob's side) are applied, where the control qubits are in state $\ket{1_l}_A\ket{0_l}_B$, and the target qubits are in state $\ket{\Phi^+_l}$.

For the zero photon case:
\begin{align}
(\ket{0_l,0_p}_A\ket{0_l,0_p}_B  ) \ket{\Phi^+_l} \ket{\Phi^+_p} \stackrel{4 \times \text{CZ}}{\rightarrow}  
(\ket{0_l,0_p}_A)  \ket{0_l,0_p}_B ) \ket{\Phi^+_l} \ket{\Phi^+_p}.
\end{align}

For the single-photon case:
\begin{align}
(\ket{1_l,0_p}_A\ket{0_l,0_p}_B , \ket{0_l,0_p}_A \ket{0_l,1_p}_B ) \ket{\Phi^+_l} \ket{\Phi^+_p} \stackrel{4 \times \text{CZ}}{\rightarrow}  
(\ket{1_l,0_p}_A\ket{0_l,0_p}_B , \ket{0_l,0_p}_A \ket{0_l,1_p}_B )  \ket{\Phi^-_l} \ket{\Phi^+_p}.
\end{align}

For the two-photon case:
\begin{align}
(\ket{1_l,1_p}_A\ket{0_l,0_p}_B , \ket{0_l,0_p}_A \ket{1_l,1_p}_B ) \ket{\Phi^+_l} \ket{\Phi^+_p} \stackrel{4 \times \text{CZ}}{\rightarrow}  
(\ket{1_l,1_p}_A\ket{0_l,0_p}_B , \ket{0_l,0_p}_A \ket{1_l,1_p}_B ) \ket{\Phi^-_l} \ket{\Phi^-_p}.
\end{align}

We would like to remove the following component from the two-photon case since there is an extra phase $\delta$ that was introduced by the STIRAP interaction:
\begin{align}
(\ket{1_l,0_p}_A\ket{1_l,0_p}_B  ) \ket{\Phi^+_l} \ket{\Phi^+_p} \stackrel{4 \times \text{CZ}}{\rightarrow}  
(\ket{1_l,0_p}_A\ket{1_l,0_p}_B ) \ket{\Phi^+_l} \ket{\Phi^+_p}.
\end{align}

Therefore if we measure the auxiliary $\ket{\Phi^\pm}$, we can distinguish between 
\begin{enumerate}
\item When both the Bell pairs are in state $\ket{\Phi^+}$, this means no photon has arrived, or the system is in the ``contaminated" two-photon state.
\item We have $\ket{\Phi^-_l}\ket{\Phi^+_p}$, which means exactly one photon has arrived.
\item We have $\ket{\Phi^-_l}\ket{\Phi^-_p}$, we have two photons in the system, where the phase component is $e^{i 2\phi}$.
\end{enumerate}

After the STIRAP interaction, the term $\sqrt{2}e^{i\phi}\ket{1,1}$ acquires an additional unwanted relative phase $e^{i\delta}$ due to coupling with the cavity, which we need to remove.

We can distinguish the zero-, single- and two-photon contributions by using parity checks on the memory qubits as well as the cavity photon state by using shared Bell states between Alice and Bob. The parity check on the cavity state can be achieved by introducing another auxiliary atom.
We use the subscripts $l$ to denote levels of an atomic system and $p$ for photonic Fock states. We assume the Bell pairs 
$\ket{\Phi^+}$ are shared between Alice and Bob, where the subscript denotes which system they are paired with. For example,
 \begin{align}
\ket{1_l}_A\ket{0_l}_B \ket{\Phi^+_l}\stackrel{2 \times \text{CZ}}{\rightarrow} \ket{1_l}_A\ket{0_l}_B \ket{\Phi^-_l}
\end{align}
 denotes that 2 CZ gates are applied, where the control qubits are $\ket{1_l}_A\ket{0_l}_B$, and the target qubits are $\ket{\Phi^+_l}$.

For the zero photon cases:
\begin{align}
&(\ket{0_l,0_p}_A\ket{0_l,0_p}_B  ) \ket{\Phi^+_l} \ket{\Phi^+_p} \stackrel{4 \times \text{CZ}}{\rightarrow}  \nn
&(\ket{0_l,0_p}_A)  \ket{1_l,1_p}_B ) \ket{\Phi^+_l} \ket{\Phi^+_p}.
\end{align}

For the single photon case:
\begin{align}
&(\ket{1_l,0_p}_A\ket{0_l,0_p}_B , \ket{0_l,0_p}_A \ket{0_l,1_p}_B ) \ket{\Phi^+_l} \ket{\Phi^+_p} \stackrel{4 \times \text{CZ}}{\rightarrow}  \nn
&(\ket{1_l,0_p}_A\ket{0_l,0_p}_B , \ket{0_l,0_p}_A \ket{0_l,1_p}_B )  \ket{\Phi^-_l} \ket{\Phi^+_p}.
\end{align}

For the two-photon cases:
\begin{align}
&(\ket{1_l,1_p}_A\ket{0_l,0_p}_B , \ket{0_l,0_p}_A \ket{1_l,1_p}_B ) \ket{\Phi^+_l} \ket{\Phi^+_p} \stackrel{4 \times \text{CZ}}{\rightarrow}  \nn
&(\ket{1_l,1_p}_A\ket{0_l,0_p}_B , \ket{0_l,0_p}_A \ket{1_l,1_p}_B ) \ket{\Phi^-_l} \ket{\Phi^-_p}.
\end{align}

We would like to remove the following component from the two-photon case since there is an extra phase $\delta$ that was introduced by the STIRAP interaction:
\begin{align}
&(\ket{1_l,0_p}_A\ket{1_l,0_p}_B  ) \ket{\Phi^+_l} \ket{\Phi^+_p} \stackrel{4 \times \text{CZ}}{\rightarrow}  \nn
&(\ket{1_l,0_p}_A\ket{1_l,0_p}_B ) \ket{\Phi^+_l} \ket{\Phi^+_p}.
\end{align}

Therefore if we measure the auxiliary $\ket{\Phi^\pm}$, we can distinguish between 
\begin{enumerate}
\item When both the Bell pairs are in state $\ket{\Phi^+}$, this means no photon has arrived, or the system is in the ``contaminated" two-photon state.
\item We have $\ket{\Phi^-_l}\ket{\Phi^+_p}$, which means exactly one photon has arrived.
\item We have $\ket{\Phi^-_l}\ket{\Phi^-_p}$, we have two photons in the system, where the phase component is $e^{i 2\phi}$.
\end{enumerate}

If we take into account 2-photon contributions, the one-photon events occur with probability 
\begin{align}
p(1,0) + p(0,1) = \frac{8 \epsilon  \left(\gamma ^2 (-\epsilon )+\epsilon +2\right)}{\left(\epsilon  \left(\gamma ^2 (-\epsilon )+\epsilon +4\right)+4\right)^2}
 \approx \epsilon - \frac{1}{2} \left(\gamma ^2+3\right) \epsilon ^2
\end{align}
which is obtained from taking the trace of $\rho^{(1)}$ in Eq.~\eqref{eq:rho1}, taking $\epsilon$ to second order.

The two-photon events occur with probability
\begin{align}
p(1,1)+p(2,0)+p(0,2)&= -\frac{4 \epsilon ^2 \left(3 \gamma ^4 \epsilon ^2-2 \gamma ^2 \left(3 \epsilon ^2+6 \epsilon -2\right)+3 (\epsilon +2)^2\right)}{((\gamma -1) \epsilon -2)^3 (\gamma  \epsilon +\epsilon +2)^3}  \nn
&\approx \frac{1}{4} \left(\gamma ^2+3\right) \epsilon ^2.
\end{align}
which is obtained from taking the trace of $\rho^{(2)}$ in Eq.~\eqref{eq:rho_two_photons}, once again, taking $\epsilon$ to second order.

The mean photon number, for the purpose of normalisation, is
\begin{align}
1\times \left(p(1,0) + p(0,1) \right) + 2\times \left(p(1,1)+p(2,0)+p(0,2) \right) = \epsilon.
\end{align}

The two-photon contributions (after STIRAP) will be


\begin{align} \label{eq:twophotons}
\rho^{(2)'}&= \left(\frac{1}{4}-\frac{\gamma ^2}{4}\right) \epsilon ^2 \ket{\psi_1}\bra{\psi_1} +
\frac{1}{4} (\gamma +1)^2 \epsilon ^2 \times \frac{1}{2}\times 
\frac{\ket{1,1}_A\ket{0,0}_B + e^{i 2\phi}\ket{0,0}_A\ket{1,1}_B}{2}\otimes h.c \nn
&+\frac{1}{4} (\gamma -1)^2 \epsilon ^2\times \frac{1}{2} \times 
\frac{\ket{1,1}_A\ket{0,0}_B + e^{i 2\phi}\ket{0,0}_A\ket{1,1}_B}{2}\otimes h.c \nn
&=\left(\frac{1}{4}-\frac{\gamma ^2}{4}\right) \epsilon ^2 
\frac{\ket{1,1}_A\ket{0,0}_B - e^{i 2\phi}\ket{0,2}}{\sqrt2}\otimes h.c +
\frac{1}{4} \left(\gamma ^2+1\right) \epsilon^2 \frac{\ket{1,1}_A\ket{0,0}_B + e^{i 2\phi}\ket{0,0}_A\ket{1,1}_B}{\sqrt{2}}\otimes h.c \nn
&=\frac{\epsilon^2}{2}\left[\left( \frac{1-\gamma ^2}{2}\right) \frac{\ket{1,1}_A\ket{0,0}_B - e^{i 2\phi}\ket{0,0}_A\ket{1,1}_B}{\sqrt2}\otimes h.c +
 \left(\frac{1+\gamma ^2}{2}\right)  \frac{\ket{1,1}_A\ket{0,0}_B + e^{i 2\phi}\ket{0,0}_A\ket{1,1}_B}{\sqrt2}\otimes h.c \right]\nn
\end{align}
The QFI's for $\gamma $ and $\phi$ for the state in Eq.~\eqref{eq:twophotons}
\begin{align}
\text{QFI}^{(2)}_\gamma &= \frac{ 2 \gamma ^2 \epsilon ^2}{(1-\gamma ^4)}, \nn
\text{QFI}^{(2)}_\phi &=  {2\epsilon^2} \gamma^4.
\end{align}

The FI's for $\gamma$ of the state in Eq.~\eqref{eq:twophotons} when measured in the local bases is
\begin{align}
\text{FI}^{(2)}_\gamma = -\frac{2 \gamma ^2 \epsilon ^2 \cos ^2(\alpha -2 \phi )}{\gamma ^4 \cos ^2(\alpha -2 \phi )-1},
\rightarrow \frac{ 2 \gamma ^2 \epsilon ^2}{(1-\gamma ^4)}
\end{align}
when we set $\alpha-2 \phi \rightarrow 0$.
As for $\phi$, 
\begin{align}
\text{FI}^{(2)}_\phi &= -\frac{2 \gamma ^4 \epsilon ^2 \sin ^2(\alpha -2 \phi )}{\gamma ^4 \cos ^2(\alpha -2 \phi )-1} \nn
&\rightarrow  2 \gamma ^4 \epsilon ^2
\end{align}
when we set $\alpha - 2\phi \rightarrow \pi/2$.

Since the one- and two-photon events are distinguishable, summing the weighted contributions, the FI of $\gamma$ per photon is
\begin{align}
\text{FI'}_\gamma &=
\frac{1}{\epsilon}\left[
\left(\epsilon - \frac{1}{2}(\gamma^2 + 3)\epsilon^2 \right)\times \frac{1}{1-\gamma^2} + \frac{ 2 \gamma ^2 \epsilon ^2}{(1-\gamma ^4)}\right] \nn
&= \frac{2+ 2\gamma^2 -3\epsilon - \epsilon\gamma^4 }{2-2 \gamma ^4}.
\end{align}

And for $\phi$:

\begin{align}\label{eq:include}
\text{FI'}_\phi &= \frac{1}{\epsilon}\left[
\left(\epsilon - \frac{1}{2}(\gamma^2 + 3)\epsilon^2 \right) \times \gamma^2+ {2\epsilon^2} \gamma^4 \right] \nn
&=\frac{1}{2} \gamma ^2 \left( 2 - 3 \epsilon \left(1-\gamma ^2\right)  \right).
\end{align}


\section{Errors associated with ignoring multiple photon contributions}

Without accounting for two-photon contributions, we use a single atom instead of two, and 
 STIRAP takes our system to the states

\begin{align}
\ket{\Psi_0^2} &\rightarrow \frac{1}{\sqrt2}(\ket{1_l,1_p}_A \ket{0_l,0_p}_B - e^{i 2\phi}\ket{0_l,0_p}_A \ket{1_l,1_p}_B) \nn
\ket{\Psi_+^2} &\rightarrow \frac{1}{2}\ket{1_l,1_p}_A \ket{0_l,0_p}_B + 
  \frac{1}{\sqrt2}e^{i(\phi + \delta)}\ket{1_l,0_p}_A\ket{1_l,0_p}_B + 
  \frac{1}{2}e^{i2\phi}\ket{0_l,0_p}_A \ket{1_l,1_p}_B \nn
\ket{\Psi_-^2} &\rightarrow \frac{1}{2}\ket{1_l,1_p}_A \ket{0_l,0_p}_B - 
  \frac{1}{\sqrt2}e^{i(\phi + \delta)}\ket{1_l,0_p}_A\ket{1_l,0_p}_B + 
  \frac{1}{2}e^{i2\phi}\ket{0_l,0_p}_A \ket{1_l,1_p}_B.
\end{align}

Performing the $2\times$ CZ gate parity check projects the state onto the odd parity subspace; the photonic Fock states part of the Hilbert space (subscripted $p$) is traced out. After the STIRAP interaction, Alice and Bob have:
\begin{align}
\ket{\Psi_0^2} &\rightarrow \ket{\psi_0^{(2)}} = 
\frac{1}{\sqrt2}(\ket{1_L,0_L}_{AB} - e^{i 2\phi}\ket{0_L,1_L}_{AB}) \nn
\ket{\Psi_+^2} &\rightarrow \ket{\psi^{(2)}_+} = \frac{1}{2}\ket{1_L,0_L}_{AB} + 
  \frac{1}{2}e^{i2\phi}\ket{0_L,1_L}_{AB} \nn
\ket{\Psi_-^2}&\rightarrow \ket{\psi^{(2)}_-} = \frac{1}{2}\ket{1_L,0_L}_{AB}  + 
  \frac{1}{2}e^{i2\phi}\ket{0_L,1_L}_{AB}
\end{align}
Note that the last two are not normalised since one of the components is projected out.

This means that the density matrix becomes
\begin{align} \label{eq:two_photon_fin}
\rho'' \approx &\left[\frac{1}{4} \left(-\gamma ^2-4 \gamma -3\right) \epsilon ^2+\frac{1}{2} (\gamma +1) \epsilon\right]
 \ket{\psi_{+,L}^\phi}\bra{\psi_{+,L}^\phi} + 
\left[\frac{1}{4} \left(-\gamma ^2+4 \gamma -3\right) \epsilon ^2+\frac{1}{2} (1-\gamma ) \epsilon \right]
\ket{\psi_{-,L}^\phi}\bra{\psi_{-,L}^\phi}  \nn
&+\left[\left(\frac{1}{4}-\frac{\gamma ^2}{4}\right) \epsilon ^2 \right] \ket{\psi_0^{(2)}}\bra{\psi_0^{(2)}} +
\frac{1}{2} \left(\gamma ^2+1\right) \epsilon ^2 \left[ \ket{\psi_+ ^{(2)}}\bra{\psi_+ ^{(2)}} \right],
\end{align}

Now, if we follow the same procedure,
when projected onto the basis $1/\sqrt2(\ket{0_L} \pm e^{i\alpha}\ket{1_L})\otimes 1/\sqrt2(\ket{0_L} \pm \ket{1_L})$, the measurement outcomes are
\begin{align}
p'(++) + p'(--) = \frac{1}{4} \epsilon  \left(\gamma ^2 \epsilon  \cos (\alpha -2 \phi )+2 \gamma  (1-2 \epsilon ) \cos (\alpha -\phi )-\left(\gamma ^2+2\right) \epsilon +2\right) \nn
p'(+-) + p'(-+) =-\frac{1}{4} \epsilon  \left(\gamma ^2 \epsilon  \cos (\alpha -2 \phi )+2 \gamma  (1-2 \epsilon ) \cos (\alpha -\phi )+\left(\gamma ^2+2\right) \epsilon -2\right).
\end{align}

The FI for the state in Eq.\eqref{eq:two_photon_fin} per photon is
\begin{align}
\text{FI}_\gamma' =
\frac{1}{4} \bigg(&\frac{4 (\gamma  \epsilon  (\cos (\alpha -2 \phi )-1)+(1-2 \epsilon ) \cos (\alpha -\phi ))^2}{\gamma ^2 \epsilon  \cos (\alpha -2 \phi )+2 \gamma  (1-2 \epsilon ) \cos (\alpha -\phi )-\left(\gamma ^2+2\right) \epsilon +2}-\nn
&\frac{4 (\gamma  \epsilon  (\cos (\alpha -2 \phi )+1)+(1-2 \epsilon ) \cos (\alpha -\phi ))^2}{\gamma ^2 \epsilon  \cos (\alpha -2 \phi )+2 \gamma  (1-2 \epsilon ) \cos (\alpha -\phi )+\left(\gamma ^2+2\right) \epsilon -2}\bigg)
\end{align}

\begin{align} \label{eq:phi_ignore}
\text{FI}_\phi' =& \frac{X'}{Y'} , \nn
 X' =& 2 \gamma ^2 \left(\left(\gamma ^2+2\right) \epsilon -2\right) (\gamma  \epsilon  \sin (\alpha -2 \phi )+(1-2 \epsilon ) \sin (\alpha -\phi ))^2. \nn
Y' = &\left(\gamma ^2 \epsilon  \cos (\alpha -2 \phi )-2 \gamma  (2 \epsilon -1) \cos (\alpha -\phi )+\gamma ^2 (-\epsilon )-2 \epsilon +2\right) \times\nn
&\left(\gamma ^2 \epsilon  \cos (\alpha -2 \phi )-2 \gamma  (2 \epsilon -1) \cos (\alpha -\phi )+\gamma ^2 \epsilon +2 \epsilon -2\right)
\end{align}

\begin{figure}[h!]
\includegraphics[trim = 0cm 0cm 0cm 0.2cm, clip, width=0.5\linewidth]{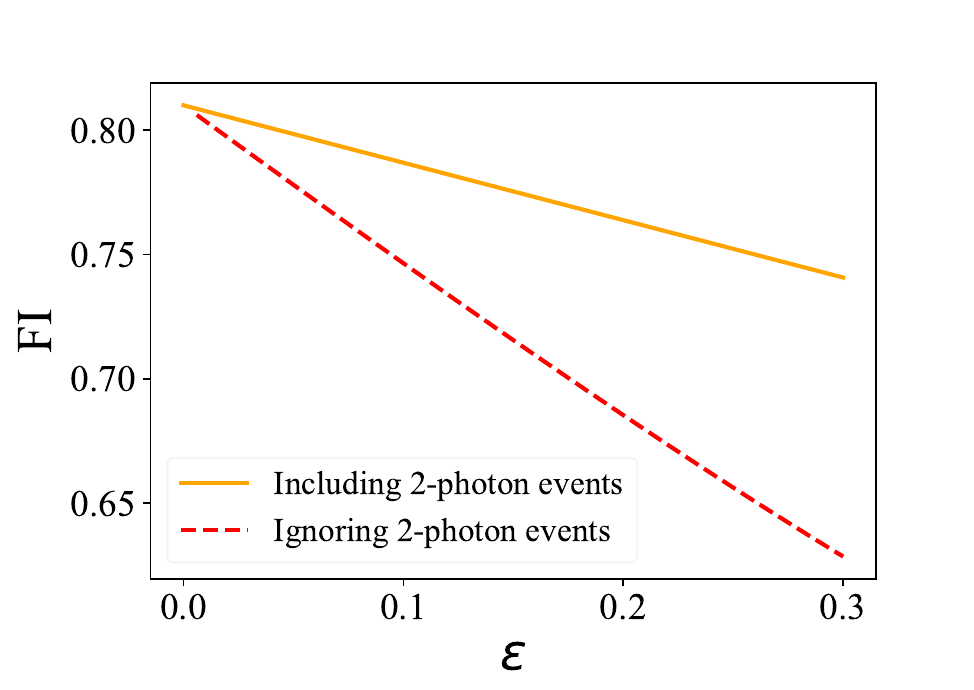} 
 \caption{\label{f:higher_ord}The FI \textit{per photon} of $\phi$ when we include (Eq.~\eqref{eq:include}) and ignore (Eq.~\eqref{eq:phi_ignore}) two-photon contributions, for $\gamma = 0.9, \phi = 0$, where $\alpha$ is optimised.}
\end{figure}

Note that the terms are now dependent on both $\alpha-\phi$ as well as $\alpha - 2\phi$, which will complicate the estimation process.

In Fig.~\ref{f:higher_ord}, for $\gamma = 0.9$, we show the FI as a function of $\epsilon$, when we include and ignore the two-photon contributions when $\alpha$ is optimised over $[0,2\pi]$. The larger $\epsilon$, the more detrimental the effect of ignoring such higher photon events.

\color{black}
\section{QEC for the depolarising channel}

The depolarising channel acting on each qubit can be written as 
\begin{align}
\mathcal{E}_\text{depol}(\rho) = (1- p) \rho + p I/2,
\end{align}
\noindent 
where 
$(1-p)$ is the probability that the transmission is noiseless, and with probability $p$ the state is replaced by the completely mixed state.

\begin{figure}[t!]
\includegraphics[trim = 0.2cm 0.0cm 0.3cm 0cm, clip, width=0.5\linewidth]{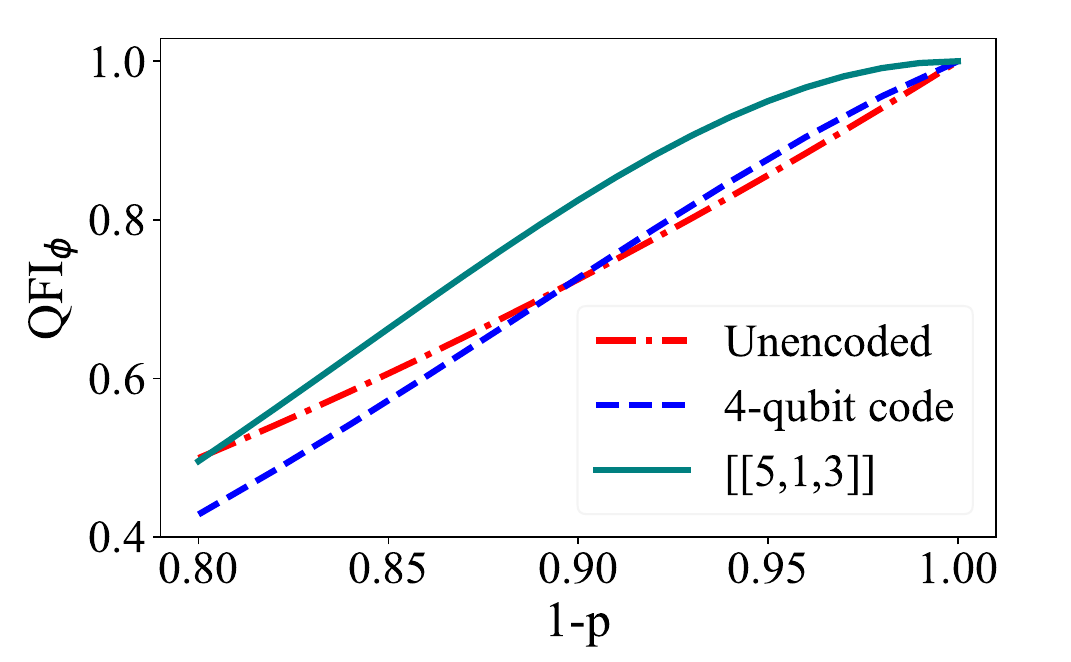} 
 \caption{\label{f:qec_phi} For $\gamma = 1$, when we use no encoding (red dotted-dashed line)
 the depolarising channel when we use the 4-qubit code \cite{LNCY97} (blue dashed line) and the [[5,1,3]] error correction code (teal solid line).}
\end{figure}

In Fig.~\ref{f:qec_phi} we show the QFI of $\phi$ per photon received as a function of the noise strength $p$ for the depolarizing channels. 
%
%
%
For the depolarising channel, the unprotected case without QEC has a QFI equal to $2(1-p)^4/(2-2p+p^2)$. Here, the [[5,1,3]] code is able to offer protection for values of $p$ up to about 20\%. This feature arises because of the favourable distance-to-length ratio of the [[5,1,3]] code.

\begin{figure}[t!]
\includegraphics[trim = 0cm 0cm 0cm 0.2cm, clip, width=0.5\linewidth]{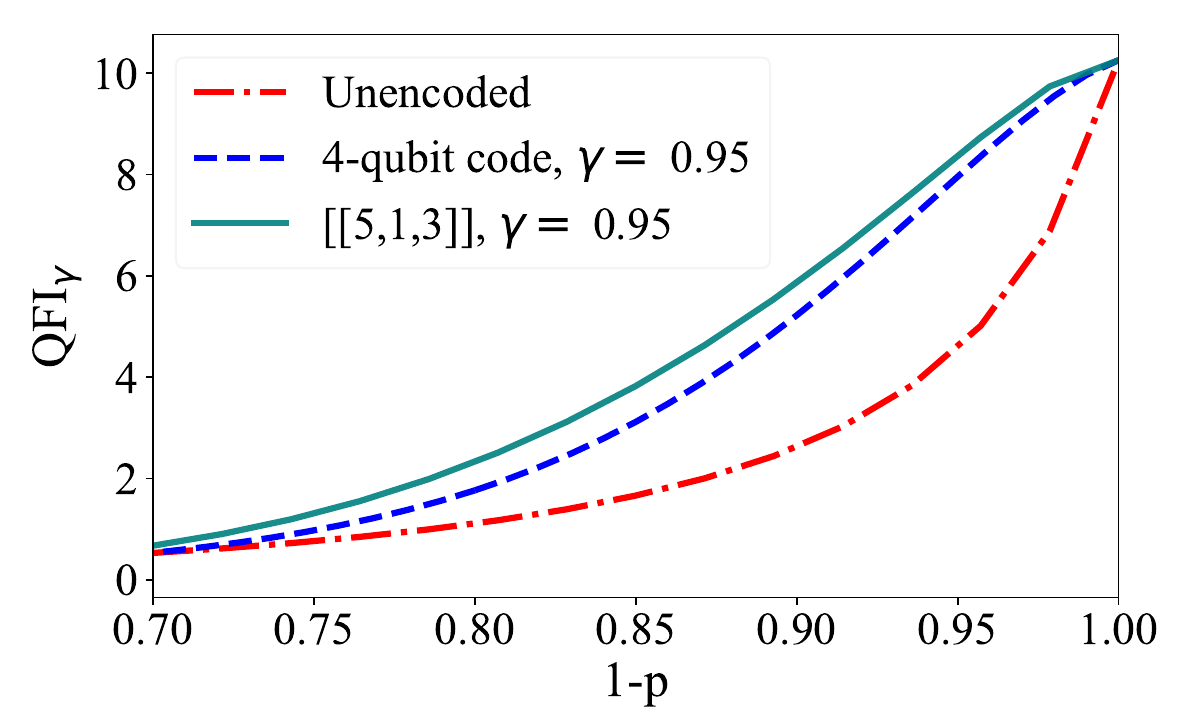} 
 \caption{\label{f:qec_gamma} For $\gamma = 0.95$,  when no encoding is used (red dotted-dashed line)
 the depolarising channel when we use the  4-qubit code (blue dashed line) and the [[5,1,3]] error correction code (teal solid line).}
\end{figure}
In Fig.~\ref{f:qec_gamma} we show the QFI of $\gamma$ per photon received. 
For the depolarising channel, we see that the four-qubit code \cite{LNCY97} with logical codewords $(|0000\rangle + |1111\rangle)/\sqrt 2$ and $(|0011\rangle + |1100\rangle)/\sqrt 2$ and the [[5,1,3]] QEC code also offers protection for a modest range of $p$.

\section{Further discussions}

The predominant type of resource required in QEC is the number of qubits. The type of QEC does indeed affects the minimum number of qubits required for the QFI to be preserved up to a given amount.

The type of QEC that one would choose to use would depend greatly on the noise model.
In certain architectures where dephasing is dominant (compared to bit flip), we can invoke QEC codes designed for biased noise thereby improving the overall performance and reducing the number of qubits needed \cite{PhysRevLett.120.050505,PRXQuantum.2.030345} as compared to surface codes \cite{Cleland12} which treat phase and bit-flip errors symmetrically.

When the quantum registers used in our protocol consist of trapped Rydberg atoms, e.g.$^{171}$Yb, errors that physically occur can be converted to erasure errors, or deletion errors if we do not keep track of which qubits are erased. In this instance, it has been recently shown that up to 98\% of the errors can be converted into erasure errors \cite{wu2022erasure}.
When the noise model introduces predominantly erasure errors, we can use $n$-qubit codes of distance $d$ for which the relative distance $d/n$ is maximised for a large $n$. There are families of quantum codes, both random \cite{ouyang2014concatenated} and explicit \cite{ALT01,CLX01,LXW09} quantum codes, for which $d/n >0$ asymptotically, with one prominent recent example being quantum low-density parity check codes \cite{panteleev2022asymptotically}.
This leads to several choices of quantum codes that one can use to protect the QFI provided the fraction of deletions is strictly less than $d/n$.


If we use random codes that almost surely attain the quantum Gilbert-Varshamov bound, this allows us to tolerate up to $18.9\%$ of erasure errors while having a robust QFI. If we are satisfied to have a lower QFI, we can use the five-qubit code, where the QFI is preserved if no more than two out of the five qubits are corrupted.
If we had not converted the physical noise model to erasure errors, the number of qubits needed to achieve a comparable performance for the QFI is potentially different. Making this statement quantitative is however a subject that we leave for future studies.

If we are unable to track which qubits are erased, then we will have a more challenging deletion errors noise model. In the absence of information about which qubits were removed, use of conventional QECs will often result in a catastrophic reduction in QFI, because a single deletion error can corrupt a large consecutive block of qubits. 
Fortunately, one can combat deletion errors using permutation-invariant quantum codes, for which deletion errors are equivalent to erasure errors \cite{ouyang2021permutation,shibayama2021permutation}.
Using $n$-qubit permutation-invariant codes, up to $O(\sqrt n)$ deletions can be corrected \cite{ouyang2021permutation}.
The shortest non-trivial permutation-invariant quantum code is the four-qubit code (2,2,1) gnu code \cite{ouyang2014permutation,nakayama2020first}, which when used in our scheme, preserves the QFI if no more than one out of the four qubits is corrupted.
A noise model that randomly deletes qubits and inserts separable states can also be corrected using permutation-invariant codes, because of the equivalence in the correctibility of deletion and separable insertion errors \cite{shibayama2021equivalence}.

In summary, the choice of QEC codes greatly affects (1) whether it is possible to make the QFI robust and (2) the number of qubits needed to make the QFI robust up to a certain degree.

In separate ongoing research work, we are also investigating the performance of QEC codes in our scheme under a particular circuit-level noise model \cite{baireuther2019neural}
, where errors are introduced probabilistically whenever quantum gates are applied. This error model can describe errors that occur in different stages of our scheme. Our ongoing work seeks to quantify the relationship between the number of qubits required to attain a particular value of the QFI and the noise strength in our error model.

\color{black}

We anticipate that by leveraging on the theory of fault-tolerant quantum computation \cite{campbell2017roads}, our scheme can achieve a high QFI even with imperfect QEC operations. 
We will need in particular fault-tolerant syndrome extraction, using for instance the method of flag qubits \cite{Chao2020.anyflag},
which only requires a modest overhead in the number of ancillary qubits.
 Namely, using any distance $d$ stabilizer code, one only needs $d+1$ additional ancilla qubits. 

Additionally, one could view the star as a source of preparing non-Clifford states, given that we have high-precision estimation of the phase and a stable source, e.g. in a satellite-based detection context.

We have mainly considered our protocol for an estimation scheme, but note that when a photon has been captured, Alice and Bob share an entangled state. This entanglement is generated entirely by the non-local properties of the single photon state \cite{PhysRevA.100.042332}. For universal quantum computation, non-Clifford states are required on top of the Clifford set \cite{campbell2017roads}. In a laboratory, we can control the phase $\phi$ by varying the position of the photon source---as opposed to applying universal phase gates---to create the non-Clifford state $(\ket{01} + e^{i\phi} \ket{10})/\sqrt 2$. Our scheme therefore can be used as an alternative approach for creating resources that can be harnessed for quantum computation.  

 If noise is below a certain threshold and we have access to additional ancillary qubits, then the theory of fault-tolerant quantum computation guarantees us that the quantum error correction procedure can be performed arbitrarily well. This in turn allows us to recover the near-optimal QFI provided that the physical noise is below the fault-tolerant threshold. The fault-tolerant resources overheads required in terms of the number of additional qubits required can be reduced at the expense of a reduction of the QFI.  Hence there is a tradeoff between available resources and the attainable QFI in the noisy (and imperfect QEC) scenario. 


\begin{thebibliography}{80}%
\makeatletter
\providecommand \@ifxundefined [1]{%
 \@ifx{#1\undefined}
}%
\providecommand \@ifnum [1]{%
 \ifnum #1\expandafter \@firstoftwo
 \else \expandafter \@secondoftwo
 \fi
}%
\providecommand \@ifx [1]{%
 \ifx #1\expandafter \@firstoftwo
 \else \expandafter \@secondoftwo
 \fi
}%
\providecommand \natexlab [1]{#1}%
\providecommand \enquote  [1]{``#1''}%
\providecommand \bibnamefont  [1]{#1}%
\providecommand \bibfnamefont [1]{#1}%
\providecommand \citenamefont [1]{#1}%
\providecommand \href@noop [0]{\@secondoftwo}%
\providecommand \href [0]{\begingroup \@sanitize@url \@href}%
\providecommand \@href[1]{\@@startlink{#1}\@@href}%
\providecommand \@@href[1]{\endgroup#1\@@endlink}%
\providecommand \@sanitize@url [0]{\catcode `\\12\catcode `\$12\catcode
  `\&12\catcode `\#12\catcode `\^12\catcode `\_12\catcode `\%12\relax}%
\providecommand \@@startlink[1]{}%
\providecommand \@@endlink[0]{}%
\providecommand \url  [0]{\begingroup\@sanitize@url \@url }%
\providecommand \@url [1]{\endgroup\@href {#1}{\urlprefix }}%
\providecommand \urlprefix  [0]{URL }%
\providecommand \Eprint [0]{\href }%
\providecommand \doibase [0]{http://dx.doi.org/}%
\providecommand \selectlanguage [0]{\@gobble}%
\providecommand \bibinfo  [0]{\@secondoftwo}%
\providecommand \bibfield  [0]{\@secondoftwo}%
\providecommand \translation [1]{[#1]}%
\providecommand \BibitemOpen [0]{}%
\providecommand \bibitemStop [0]{}%
\providecommand \bibitemNoStop [0]{.\EOS\space}%
\providecommand \EOS [0]{\spacefactor3000\relax}%
\providecommand \BibitemShut  [1]{\csname bibitem#1\endcsname}%
\let\auto@bib@innerbib\@empty
\bibitem [{\citenamefont {Giovannetti}\ \emph {et~al.}(2006)\citenamefont
  {Giovannetti}, \citenamefont {Lloyd},\ and\ \citenamefont
  {Maccone}}]{giovannetti2006quantum}%
  \BibitemOpen
  \bibfield  {author} {\bibinfo {author} {\bibfnamefont {V.}~\bibnamefont
  {Giovannetti}}, \bibinfo {author} {\bibfnamefont {S.}~\bibnamefont {Lloyd}},
  \ and\ \bibinfo {author} {\bibfnamefont {L.}~\bibnamefont {Maccone}},\
  }\href@noop {} {\bibfield  {journal} {\bibinfo  {journal} {Physical review
  letters}\ }\textbf {\bibinfo {volume} {96}},\ \bibinfo {pages} {010401}
  (\bibinfo {year} {2006})}\BibitemShut {NoStop}%
\bibitem [{\citenamefont {Giovannetti}\ \emph {et~al.}(2011)\citenamefont
  {Giovannetti}, \citenamefont {Lloyd},\ and\ \citenamefont
  {Maccone}}]{giovannetti2011advances}%
  \BibitemOpen
  \bibfield  {author} {\bibinfo {author} {\bibfnamefont {V.}~\bibnamefont
  {Giovannetti}}, \bibinfo {author} {\bibfnamefont {S.}~\bibnamefont {Lloyd}},
  \ and\ \bibinfo {author} {\bibfnamefont {L.}~\bibnamefont {Maccone}},\
  }\href@noop {} {\bibfield  {journal} {\bibinfo  {journal} {Nature photonics}\
  }\textbf {\bibinfo {volume} {5}},\ \bibinfo {pages} {222} (\bibinfo {year}
  {2011})}\BibitemShut {NoStop}%
\bibitem [{\citenamefont {Lichtman}\ and\ \citenamefont
  {Conchello}(2005)}]{lichtman2005fluorescence}%
  \BibitemOpen
  \bibfield  {author} {\bibinfo {author} {\bibfnamefont {J.~W.}\ \bibnamefont
  {Lichtman}}\ and\ \bibinfo {author} {\bibfnamefont {J.-A.}\ \bibnamefont
  {Conchello}},\ }\href@noop {} {\bibfield  {journal} {\bibinfo  {journal}
  {Nature methods}\ }\textbf {\bibinfo {volume} {2}},\ \bibinfo {pages} {910}
  (\bibinfo {year} {2005})}\BibitemShut {NoStop}%
\bibitem [{\citenamefont {Casacio}\ \emph {et~al.}(2021)\citenamefont
  {Casacio}, \citenamefont {Madsen}, \citenamefont {Terrasson}, \citenamefont
  {Waleed}, \citenamefont {Barnscheidt}, \citenamefont {Hage}, \citenamefont
  {Taylor},\ and\ \citenamefont {Bowen}}]{casacio2021quantum}%
  \BibitemOpen
  \bibfield  {author} {\bibinfo {author} {\bibfnamefont {C.~A.}\ \bibnamefont
  {Casacio}}, \bibinfo {author} {\bibfnamefont {L.~S.}\ \bibnamefont {Madsen}},
  \bibinfo {author} {\bibfnamefont {A.}~\bibnamefont {Terrasson}}, \bibinfo
  {author} {\bibfnamefont {M.}~\bibnamefont {Waleed}}, \bibinfo {author}
  {\bibfnamefont {K.}~\bibnamefont {Barnscheidt}}, \bibinfo {author}
  {\bibfnamefont {B.}~\bibnamefont {Hage}}, \bibinfo {author} {\bibfnamefont
  {M.~A.}\ \bibnamefont {Taylor}}, \ and\ \bibinfo {author} {\bibfnamefont
  {W.~P.}\ \bibnamefont {Bowen}},\ }\href@noop {} {\bibfield  {journal}
  {\bibinfo  {journal} {Nature}\ }\textbf {\bibinfo {volume} {594}},\ \bibinfo
  {pages} {201} (\bibinfo {year} {2021})}\BibitemShut {NoStop}%
\bibitem [{\citenamefont {Khabiboulline}\ \emph
  {et~al.}(2019{\natexlab{a}})\citenamefont {Khabiboulline}, \citenamefont
  {Borregaard}, \citenamefont {De~Greve},\ and\ \citenamefont
  {Lukin}}]{PhysRevLett.123.070504}%
  \BibitemOpen
  \bibfield  {author} {\bibinfo {author} {\bibfnamefont {E.~T.}\ \bibnamefont
  {Khabiboulline}}, \bibinfo {author} {\bibfnamefont {J.}~\bibnamefont
  {Borregaard}}, \bibinfo {author} {\bibfnamefont {K.}~\bibnamefont
  {De~Greve}}, \ and\ \bibinfo {author} {\bibfnamefont {M.~D.}\ \bibnamefont
  {Lukin}},\ }\href {\doibase 10.1103/PhysRevLett.123.070504} {\bibfield
  {journal} {\bibinfo  {journal} {Phys. Rev. Lett.}\ }\textbf {\bibinfo
  {volume} {123}},\ \bibinfo {pages} {070504} (\bibinfo {year}
  {2019}{\natexlab{a}})}\BibitemShut {NoStop}%
\bibitem [{\citenamefont {Terhal}(2015)}]{RevModPhys.87.307}%
  \BibitemOpen
  \bibfield  {author} {\bibinfo {author} {\bibfnamefont {B.~M.}\ \bibnamefont
  {Terhal}},\ }\href {\doibase 10.1103/RevModPhys.87.307} {\bibfield  {journal}
  {\bibinfo  {journal} {Rev. Mod. Phys.}\ }\textbf {\bibinfo {volume} {87}},\
  \bibinfo {pages} {307} (\bibinfo {year} {2015})}\BibitemShut {NoStop}%
\bibitem [{\citenamefont {Zhou}\ \emph {et~al.}(2018)\citenamefont {Zhou},
  \citenamefont {Zhang}, \citenamefont {Preskill},\ and\ \citenamefont
  {Jiang}}]{zhou2018achieving}%
  \BibitemOpen
  \bibfield  {author} {\bibinfo {author} {\bibfnamefont {S.}~\bibnamefont
  {Zhou}}, \bibinfo {author} {\bibfnamefont {M.}~\bibnamefont {Zhang}},
  \bibinfo {author} {\bibfnamefont {J.}~\bibnamefont {Preskill}}, \ and\
  \bibinfo {author} {\bibfnamefont {L.}~\bibnamefont {Jiang}},\ }\href@noop {}
  {\bibfield  {journal} {\bibinfo  {journal} {Nature communications}\ }\textbf
  {\bibinfo {volume} {9}},\ \bibinfo {pages} {1} (\bibinfo {year}
  {2018})}\BibitemShut {NoStop}%
\bibitem [{\citenamefont {Demkowicz-Dobrza\ifmmode~\acute{n}\else
  \'{n}\fi{}ski}\ \emph {et~al.}(2017)\citenamefont
  {Demkowicz-Dobrza\ifmmode~\acute{n}\else \'{n}\fi{}ski}, \citenamefont
  {Czajkowski},\ and\ \citenamefont {Sekatski}}]{PhysRevX.7.041009}%
  \BibitemOpen
  \bibfield  {author} {\bibinfo {author} {\bibfnamefont {R.}~\bibnamefont
  {Demkowicz-Dobrza\ifmmode~\acute{n}\else \'{n}\fi{}ski}}, \bibinfo {author}
  {\bibfnamefont {J.}~\bibnamefont {Czajkowski}}, \ and\ \bibinfo {author}
  {\bibfnamefont {P.}~\bibnamefont {Sekatski}},\ }\href {\doibase
  10.1103/PhysRevX.7.041009} {\bibfield  {journal} {\bibinfo  {journal} {Phys.
  Rev. X}\ }\textbf {\bibinfo {volume} {7}},\ \bibinfo {pages} {041009}
  (\bibinfo {year} {2017})}\BibitemShut {NoStop}%
\bibitem [{\citenamefont {Huang}\ \emph {et~al.}(2016)\citenamefont {Huang},
  \citenamefont {Macchiavello},\ and\ \citenamefont
  {Maccone}}]{PhysRevA.94.012101}%
  \BibitemOpen
  \bibfield  {author} {\bibinfo {author} {\bibfnamefont {Z.}~\bibnamefont
  {Huang}}, \bibinfo {author} {\bibfnamefont {C.}~\bibnamefont {Macchiavello}},
  \ and\ \bibinfo {author} {\bibfnamefont {L.}~\bibnamefont {Maccone}},\ }\href
  {\doibase 10.1103/PhysRevA.94.012101} {\bibfield  {journal} {\bibinfo
  {journal} {Phys. Rev. A}\ }\textbf {\bibinfo {volume} {94}},\ \bibinfo
  {pages} {012101} (\bibinfo {year} {2016})}\BibitemShut {NoStop}%
\bibitem [{\citenamefont {Shettell}\ \emph {et~al.}(2021)\citenamefont
  {Shettell}, \citenamefont {Munro}, \citenamefont {Markham},\ and\
  \citenamefont {Nemoto}}]{shettell2021practical}%
  \BibitemOpen
  \bibfield  {author} {\bibinfo {author} {\bibfnamefont {N.}~\bibnamefont
  {Shettell}}, \bibinfo {author} {\bibfnamefont {W.~J.}\ \bibnamefont {Munro}},
  \bibinfo {author} {\bibfnamefont {D.}~\bibnamefont {Markham}}, \ and\
  \bibinfo {author} {\bibfnamefont {K.}~\bibnamefont {Nemoto}},\ }\href@noop {}
  {\bibfield  {journal} {\bibinfo  {journal} {New Journal of Physics}\ }\textbf
  {\bibinfo {volume} {23}},\ \bibinfo {pages} {043038} (\bibinfo {year}
  {2021})}\BibitemShut {NoStop}%
\bibitem [{\citenamefont {Kessler}\ \emph {et~al.}(2014)\citenamefont
  {Kessler}, \citenamefont {Lovchinsky}, \citenamefont {Sushkov},\ and\
  \citenamefont {Lukin}}]{PhysRevLett.112.150802}%
  \BibitemOpen
  \bibfield  {author} {\bibinfo {author} {\bibfnamefont {E.~M.}\ \bibnamefont
  {Kessler}}, \bibinfo {author} {\bibfnamefont {I.}~\bibnamefont {Lovchinsky}},
  \bibinfo {author} {\bibfnamefont {A.~O.}\ \bibnamefont {Sushkov}}, \ and\
  \bibinfo {author} {\bibfnamefont {M.~D.}\ \bibnamefont {Lukin}},\ }\href
  {\doibase 10.1103/PhysRevLett.112.150802} {\bibfield  {journal} {\bibinfo
  {journal} {Phys. Rev. Lett.}\ }\textbf {\bibinfo {volume} {112}},\ \bibinfo
  {pages} {150802} (\bibinfo {year} {2014})}\BibitemShut {NoStop}%
\bibitem [{\citenamefont {D\"ur}\ \emph {et~al.}(2014)\citenamefont {D\"ur},
  \citenamefont {Skotiniotis}, \citenamefont {Fr\"owis},\ and\ \citenamefont
  {Kraus}}]{PhysRevLett.112.080801}%
  \BibitemOpen
  \bibfield  {author} {\bibinfo {author} {\bibfnamefont {W.}~\bibnamefont
  {D\"ur}}, \bibinfo {author} {\bibfnamefont {M.}~\bibnamefont {Skotiniotis}},
  \bibinfo {author} {\bibfnamefont {F.}~\bibnamefont {Fr\"owis}}, \ and\
  \bibinfo {author} {\bibfnamefont {B.}~\bibnamefont {Kraus}},\ }\href
  {\doibase 10.1103/PhysRevLett.112.080801} {\bibfield  {journal} {\bibinfo
  {journal} {Phys. Rev. Lett.}\ }\textbf {\bibinfo {volume} {112}},\ \bibinfo
  {pages} {080801} (\bibinfo {year} {2014})}\BibitemShut {NoStop}%
\bibitem [{\citenamefont {Unden}\ \emph {et~al.}(2016)\citenamefont {Unden},
  \citenamefont {Balasubramanian}, \citenamefont {Louzon}, \citenamefont
  {Vinkler}, \citenamefont {Plenio}, \citenamefont {Markham}, \citenamefont
  {Twitchen}, \citenamefont {Stacey}, \citenamefont {Lovchinsky}, \citenamefont
  {Sushkov}, \citenamefont {Lukin}, \citenamefont {Retzker}, \citenamefont
  {Naydenov}, \citenamefont {McGuinness},\ and\ \citenamefont
  {Jelezko}}]{PhysRevLett.116.230502}%
  \BibitemOpen
  \bibfield  {author} {\bibinfo {author} {\bibfnamefont {T.}~\bibnamefont
  {Unden}}, \bibinfo {author} {\bibfnamefont {P.}~\bibnamefont
  {Balasubramanian}}, \bibinfo {author} {\bibfnamefont {D.}~\bibnamefont
  {Louzon}}, \bibinfo {author} {\bibfnamefont {Y.}~\bibnamefont {Vinkler}},
  \bibinfo {author} {\bibfnamefont {M.~B.}\ \bibnamefont {Plenio}}, \bibinfo
  {author} {\bibfnamefont {M.}~\bibnamefont {Markham}}, \bibinfo {author}
  {\bibfnamefont {D.}~\bibnamefont {Twitchen}}, \bibinfo {author}
  {\bibfnamefont {A.}~\bibnamefont {Stacey}}, \bibinfo {author} {\bibfnamefont
  {I.}~\bibnamefont {Lovchinsky}}, \bibinfo {author} {\bibfnamefont {A.~O.}\
  \bibnamefont {Sushkov}}, \bibinfo {author} {\bibfnamefont {M.~D.}\
  \bibnamefont {Lukin}}, \bibinfo {author} {\bibfnamefont {A.}~\bibnamefont
  {Retzker}}, \bibinfo {author} {\bibfnamefont {B.}~\bibnamefont {Naydenov}},
  \bibinfo {author} {\bibfnamefont {L.~P.}\ \bibnamefont {McGuinness}}, \ and\
  \bibinfo {author} {\bibfnamefont {F.}~\bibnamefont {Jelezko}},\ }\href
  {\doibase 10.1103/PhysRevLett.116.230502} {\bibfield  {journal} {\bibinfo
  {journal} {Phys. Rev. Lett.}\ }\textbf {\bibinfo {volume} {116}},\ \bibinfo
  {pages} {230502} (\bibinfo {year} {2016})}\BibitemShut {NoStop}%
\bibitem [{\citenamefont {Layden}\ \emph {et~al.}(2019)\citenamefont {Layden},
  \citenamefont {Zhou}, \citenamefont {Cappellaro},\ and\ \citenamefont
  {Jiang}}]{PhysRevLett.122.040502}%
  \BibitemOpen
  \bibfield  {author} {\bibinfo {author} {\bibfnamefont {D.}~\bibnamefont
  {Layden}}, \bibinfo {author} {\bibfnamefont {S.}~\bibnamefont {Zhou}},
  \bibinfo {author} {\bibfnamefont {P.}~\bibnamefont {Cappellaro}}, \ and\
  \bibinfo {author} {\bibfnamefont {L.}~\bibnamefont {Jiang}},\ }\href
  {\doibase 10.1103/PhysRevLett.122.040502} {\bibfield  {journal} {\bibinfo
  {journal} {Phys. Rev. Lett.}\ }\textbf {\bibinfo {volume} {122}},\ \bibinfo
  {pages} {040502} (\bibinfo {year} {2019})}\BibitemShut {NoStop}%
\bibitem [{\citenamefont {Ouyang}\ \emph {et~al.}(2022)\citenamefont {Ouyang},
  \citenamefont {Shettell},\ and\ \citenamefont {Markham}}]{ouyang2019robust}%
  \BibitemOpen
  \bibfield  {author} {\bibinfo {author} {\bibfnamefont {Y.}~\bibnamefont
  {Ouyang}}, \bibinfo {author} {\bibfnamefont {N.}~\bibnamefont {Shettell}}, \
  and\ \bibinfo {author} {\bibfnamefont {D.}~\bibnamefont {Markham}},\ }\href
  {\doibase 10.1109/TIT.2021.3132634} {\bibfield  {journal} {\bibinfo
  {journal} {IEEE Transactions on Information Theory}\ }\textbf {\bibinfo
  {volume} {68}},\ \bibinfo {pages} {1809} (\bibinfo {year}
  {2022})}\BibitemShut {NoStop}%
\bibitem [{\citenamefont {Ouyang}\ and\ \citenamefont
  {Rengaswamy}(2020)}]{ouyang2020weight}%
  \BibitemOpen
  \bibfield  {author} {\bibinfo {author} {\bibfnamefont {Y.}~\bibnamefont
  {Ouyang}}\ and\ \bibinfo {author} {\bibfnamefont {N.}~\bibnamefont
  {Rengaswamy}},\ }\href@noop {} {\bibfield  {journal} {\bibinfo  {journal}
  {arXiv preprint arXiv:2007.02859}\ } (\bibinfo {year} {2020})}\BibitemShut
  {NoStop}%
\bibitem [{\citenamefont {Vitanov}\ \emph {et~al.}(2017)\citenamefont
  {Vitanov}, \citenamefont {Rangelov}, \citenamefont {Shore},\ and\
  \citenamefont {Bergmann}}]{RevModPhys.89.015006}%
  \BibitemOpen
  \bibfield  {author} {\bibinfo {author} {\bibfnamefont {N.~V.}\ \bibnamefont
  {Vitanov}}, \bibinfo {author} {\bibfnamefont {A.~A.}\ \bibnamefont
  {Rangelov}}, \bibinfo {author} {\bibfnamefont {B.~W.}\ \bibnamefont {Shore}},
  \ and\ \bibinfo {author} {\bibfnamefont {K.}~\bibnamefont {Bergmann}},\
  }\href {\doibase 10.1103/RevModPhys.89.015006} {\bibfield  {journal}
  {\bibinfo  {journal} {Rev. Mod. Phys.}\ }\textbf {\bibinfo {volume} {89}},\
  \bibinfo {pages} {015006} (\bibinfo {year} {2017})}\BibitemShut {NoStop}%
\bibitem [{\citenamefont {Gottesman}\ \emph {et~al.}(2012)\citenamefont
  {Gottesman}, \citenamefont {Jennewein},\ and\ \citenamefont
  {Croke}}]{PhysRevLett.109.070503}%
  \BibitemOpen
  \bibfield  {author} {\bibinfo {author} {\bibfnamefont {D.}~\bibnamefont
  {Gottesman}}, \bibinfo {author} {\bibfnamefont {T.}~\bibnamefont
  {Jennewein}}, \ and\ \bibinfo {author} {\bibfnamefont {S.}~\bibnamefont
  {Croke}},\ }\href {\doibase 10.1103/PhysRevLett.109.070503} {\bibfield
  {journal} {\bibinfo  {journal} {Phys. Rev. Lett.}\ }\textbf {\bibinfo
  {volume} {109}},\ \bibinfo {pages} {070503} (\bibinfo {year}
  {2012})}\BibitemShut {NoStop}%
\bibitem [{\citenamefont {Mandel}\ and\ \citenamefont
  {Wolf}(1995)}]{mandel1995optical}%
  \BibitemOpen
  \bibfield  {author} {\bibinfo {author} {\bibfnamefont {L.}~\bibnamefont
  {Mandel}}\ and\ \bibinfo {author} {\bibfnamefont {E.}~\bibnamefont {Wolf}},\
  }\href@noop {} {\emph {\bibinfo {title} {Optical coherence and quantum
  optics}}}\ (\bibinfo  {publisher} {Cambridge university press},\ \bibinfo
  {year} {1995})\BibitemShut {NoStop}%
\bibitem [{\citenamefont {Pearce}\ \emph {et~al.}(2017)\citenamefont {Pearce},
  \citenamefont {Campbell},\ and\ \citenamefont {Kok}}]{pearce2017optimal}%
  \BibitemOpen
  \bibfield  {author} {\bibinfo {author} {\bibfnamefont {M.~E.}\ \bibnamefont
  {Pearce}}, \bibinfo {author} {\bibfnamefont {E.~T.}\ \bibnamefont
  {Campbell}}, \ and\ \bibinfo {author} {\bibfnamefont {P.}~\bibnamefont
  {Kok}},\ }\href@noop {} {\bibfield  {journal} {\bibinfo  {journal} {Quantum}\
  }\textbf {\bibinfo {volume} {1}},\ \bibinfo {pages} {21} (\bibinfo {year}
  {2017})}\BibitemShut {NoStop}%
\bibitem [{\citenamefont {Braunstein}\ and\ \citenamefont
  {Caves}(1994)}]{caves}%
  \BibitemOpen
  \bibfield  {author} {\bibinfo {author} {\bibfnamefont {S.~L.}\ \bibnamefont
  {Braunstein}}\ and\ \bibinfo {author} {\bibfnamefont {C.~M.}\ \bibnamefont
  {Caves}},\ }\href {\doibase 10.1103/PhysRevLett.72.3439} {\bibfield
  {journal} {\bibinfo  {journal} {Phys. Rev. Lett.}\ }\textbf {\bibinfo
  {volume} {72}},\ \bibinfo {pages} {3439} (\bibinfo {year}
  {1994})}\BibitemShut {NoStop}%
\bibitem [{\citenamefont {Afnan}\ \emph {et~al.}(1996)\citenamefont {Afnan},
  \citenamefont {Banerjee}, \citenamefont {Braunstein}, \citenamefont {Brevik},
  \citenamefont {Caves}, \citenamefont {Chakraborty}, \citenamefont
  {Fischbach}, \citenamefont {Lindblom}, \citenamefont {Milburn}, \citenamefont
  {Odintsov} \emph {et~al.}}]{caves1}%
  \BibitemOpen
  \bibfield  {author} {\bibinfo {author} {\bibfnamefont {I.}~\bibnamefont
  {Afnan}}, \bibinfo {author} {\bibfnamefont {R.}~\bibnamefont {Banerjee}},
  \bibinfo {author} {\bibfnamefont {S.~L.}\ \bibnamefont {Braunstein}},
  \bibinfo {author} {\bibfnamefont {I.}~\bibnamefont {Brevik}}, \bibinfo
  {author} {\bibfnamefont {C.~M.}\ \bibnamefont {Caves}}, \bibinfo {author}
  {\bibfnamefont {B.}~\bibnamefont {Chakraborty}}, \bibinfo {author}
  {\bibfnamefont {E.}~\bibnamefont {Fischbach}}, \bibinfo {author}
  {\bibfnamefont {L.}~\bibnamefont {Lindblom}}, \bibinfo {author}
  {\bibfnamefont {G.}~\bibnamefont {Milburn}}, \bibinfo {author} {\bibfnamefont
  {S.}~\bibnamefont {Odintsov}},  \emph {et~al.},\ }\href@noop {} {\bibfield
  {journal} {\bibinfo  {journal} {Ann. Phys.}\ }\textbf {\bibinfo {volume}
  {247}},\ \bibinfo {pages} {447} (\bibinfo {year} {1996})}\BibitemShut
  {NoStop}%
\bibitem [{\citenamefont {Paris}(2009)}]{paris2009quantum}%
  \BibitemOpen
  \bibfield  {author} {\bibinfo {author} {\bibfnamefont {M.~G.}\ \bibnamefont
  {Paris}},\ }\href@noop {} {\bibfield  {journal} {\bibinfo  {journal}
  {International Journal of Quantum Information}\ }\textbf {\bibinfo {volume}
  {7}},\ \bibinfo {pages} {125} (\bibinfo {year} {2009})}\BibitemShut {NoStop}%
\bibitem [{\citenamefont {Ragy}\ \emph {et~al.}(2016)\citenamefont {Ragy},
  \citenamefont {Jarzyna},\ and\ \citenamefont
  {Demkowicz-Dobrza\ifmmode~\acute{n}\else
  \'{n}\fi{}ski}}]{PhysRevA.94.052108}%
  \BibitemOpen
  \bibfield  {author} {\bibinfo {author} {\bibfnamefont {S.}~\bibnamefont
  {Ragy}}, \bibinfo {author} {\bibfnamefont {M.}~\bibnamefont {Jarzyna}}, \
  and\ \bibinfo {author} {\bibfnamefont {R.}~\bibnamefont
  {Demkowicz-Dobrza\ifmmode~\acute{n}\else \'{n}\fi{}ski}},\ }\href {\doibase
  10.1103/PhysRevA.94.052108} {\bibfield  {journal} {\bibinfo  {journal} {Phys.
  Rev. A}\ }\textbf {\bibinfo {volume} {94}},\ \bibinfo {pages} {052108}
  (\bibinfo {year} {2016})}\BibitemShut {NoStop}%
\bibitem [{\citenamefont {Lupo}\ \emph {et~al.}(2020)\citenamefont {Lupo},
  \citenamefont {Huang},\ and\ \citenamefont {Kok}}]{PhysRevLett.124.080503}%
  \BibitemOpen
  \bibfield  {author} {\bibinfo {author} {\bibfnamefont {C.}~\bibnamefont
  {Lupo}}, \bibinfo {author} {\bibfnamefont {Z.}~\bibnamefont {Huang}}, \ and\
  \bibinfo {author} {\bibfnamefont {P.}~\bibnamefont {Kok}},\ }\href {\doibase
  10.1103/PhysRevLett.124.080503} {\bibfield  {journal} {\bibinfo  {journal}
  {Phys. Rev. Lett.}\ }\textbf {\bibinfo {volume} {124}},\ \bibinfo {pages}
  {080503} (\bibinfo {year} {2020})}\BibitemShut {NoStop}%
\bibitem [{\citenamefont {Berry}\ \emph {et~al.}(2001)\citenamefont {Berry},
  \citenamefont {Wiseman},\ and\ \citenamefont {Breslin}}]{PhysRevA.63.053804}%
  \BibitemOpen
  \bibfield  {author} {\bibinfo {author} {\bibfnamefont {D.~W.}\ \bibnamefont
  {Berry}}, \bibinfo {author} {\bibfnamefont {H.~M.}\ \bibnamefont {Wiseman}},
  \ and\ \bibinfo {author} {\bibfnamefont {J.~K.}\ \bibnamefont {Breslin}},\
  }\href {\doibase 10.1103/PhysRevA.63.053804} {\bibfield  {journal} {\bibinfo
  {journal} {Phys. Rev. A}\ }\textbf {\bibinfo {volume} {63}},\ \bibinfo
  {pages} {053804} (\bibinfo {year} {2001})}\BibitemShut {NoStop}%
\bibitem [{\citenamefont {Berry}\ and\ \citenamefont
  {Wiseman}(2000)}]{PhysRevLett.85.5098}%
  \BibitemOpen
  \bibfield  {author} {\bibinfo {author} {\bibfnamefont {D.~W.}\ \bibnamefont
  {Berry}}\ and\ \bibinfo {author} {\bibfnamefont {H.~M.}\ \bibnamefont
  {Wiseman}},\ }\href {\doibase 10.1103/PhysRevLett.85.5098} {\bibfield
  {journal} {\bibinfo  {journal} {Phys. Rev. Lett.}\ }\textbf {\bibinfo
  {volume} {85}},\ \bibinfo {pages} {5098} (\bibinfo {year}
  {2000})}\BibitemShut {NoStop}%
\bibitem [{\citenamefont {Huang}\ \emph {et~al.}(2017)\citenamefont {Huang},
  \citenamefont {Motes}, \citenamefont {Anisimov}, \citenamefont {Dowling},\
  and\ \citenamefont {Berry}}]{PhysRevA.95.053837}%
  \BibitemOpen
  \bibfield  {author} {\bibinfo {author} {\bibfnamefont {Z.}~\bibnamefont
  {Huang}}, \bibinfo {author} {\bibfnamefont {K.~R.}\ \bibnamefont {Motes}},
  \bibinfo {author} {\bibfnamefont {P.~M.}\ \bibnamefont {Anisimov}}, \bibinfo
  {author} {\bibfnamefont {J.~P.}\ \bibnamefont {Dowling}}, \ and\ \bibinfo
  {author} {\bibfnamefont {D.~W.}\ \bibnamefont {Berry}},\ }\href {\doibase
  10.1103/PhysRevA.95.053837} {\bibfield  {journal} {\bibinfo  {journal} {Phys.
  Rev. A}\ }\textbf {\bibinfo {volume} {95}},\ \bibinfo {pages} {053837}
  (\bibinfo {year} {2017})}\BibitemShut {NoStop}%
\bibitem [{\citenamefont {Jarzyna}\ and\ \citenamefont
  {Demkowicz-Dobrza\ifmmode~\acute{n}\else
  \'{n}\fi{}ski}(2013)}]{PhysRevLett.110.240405}%
  \BibitemOpen
  \bibfield  {author} {\bibinfo {author} {\bibfnamefont {M.}~\bibnamefont
  {Jarzyna}}\ and\ \bibinfo {author} {\bibfnamefont {R.}~\bibnamefont
  {Demkowicz-Dobrza\ifmmode~\acute{n}\else \'{n}\fi{}ski}},\ }\href {\doibase
  10.1103/PhysRevLett.110.240405} {\bibfield  {journal} {\bibinfo  {journal}
  {Phys. Rev. Lett.}\ }\textbf {\bibinfo {volume} {110}},\ \bibinfo {pages}
  {240405} (\bibinfo {year} {2013})}\BibitemShut {NoStop}%
\bibitem [{\citenamefont {Khabiboulline}\ \emph
  {et~al.}(2019{\natexlab{b}})\citenamefont {Khabiboulline}, \citenamefont
  {Borregaard}, \citenamefont {De~Greve},\ and\ \citenamefont
  {Lukin}}]{PhysRevA.100.022316}%
  \BibitemOpen
  \bibfield  {author} {\bibinfo {author} {\bibfnamefont {E.~T.}\ \bibnamefont
  {Khabiboulline}}, \bibinfo {author} {\bibfnamefont {J.}~\bibnamefont
  {Borregaard}}, \bibinfo {author} {\bibfnamefont {K.}~\bibnamefont
  {De~Greve}}, \ and\ \bibinfo {author} {\bibfnamefont {M.~D.}\ \bibnamefont
  {Lukin}},\ }\href {\doibase 10.1103/PhysRevA.100.022316} {\bibfield
  {journal} {\bibinfo  {journal} {Phys. Rev. A}\ }\textbf {\bibinfo {volume}
  {100}},\ \bibinfo {pages} {022316} (\bibinfo {year}
  {2019}{\natexlab{b}})}\BibitemShut {NoStop}%
\bibitem [{\citenamefont {Vasilev}\ \emph {et~al.}(2009)\citenamefont
  {Vasilev}, \citenamefont {Kuhn},\ and\ \citenamefont
  {Vitanov}}]{PhysRevA.80.013417}%
  \BibitemOpen
  \bibfield  {author} {\bibinfo {author} {\bibfnamefont {G.~S.}\ \bibnamefont
  {Vasilev}}, \bibinfo {author} {\bibfnamefont {A.}~\bibnamefont {Kuhn}}, \
  and\ \bibinfo {author} {\bibfnamefont {N.~V.}\ \bibnamefont {Vitanov}},\
  }\href {\doibase 10.1103/PhysRevA.80.013417} {\bibfield  {journal} {\bibinfo
  {journal} {Phys. Rev. A}\ }\textbf {\bibinfo {volume} {80}},\ \bibinfo
  {pages} {013417} (\bibinfo {year} {2009})}\BibitemShut {NoStop}%
\bibitem [{\citenamefont {Shore}(2017)}]{shore2017picturing}%
  \BibitemOpen
  \bibfield  {author} {\bibinfo {author} {\bibfnamefont {B.~W.}\ \bibnamefont
  {Shore}},\ }\href@noop {} {\bibfield  {journal} {\bibinfo  {journal}
  {Advances in Optics and Photonics}\ }\textbf {\bibinfo {volume} {9}},\
  \bibinfo {pages} {563} (\bibinfo {year} {2017})}\BibitemShut {NoStop}%
\bibitem [{\citenamefont {Cubel}\ \emph {et~al.}(2005)\citenamefont {Cubel},
  \citenamefont {Teo}, \citenamefont {Malinovsky}, \citenamefont {Guest},
  \citenamefont {Reinhard}, \citenamefont {Knuffman}, \citenamefont {Berman},\
  and\ \citenamefont {Raithel}}]{PhysRevA.72.023405}%
  \BibitemOpen
  \bibfield  {author} {\bibinfo {author} {\bibfnamefont {T.}~\bibnamefont
  {Cubel}}, \bibinfo {author} {\bibfnamefont {B.~K.}\ \bibnamefont {Teo}},
  \bibinfo {author} {\bibfnamefont {V.~S.}\ \bibnamefont {Malinovsky}},
  \bibinfo {author} {\bibfnamefont {J.~R.}\ \bibnamefont {Guest}}, \bibinfo
  {author} {\bibfnamefont {A.}~\bibnamefont {Reinhard}}, \bibinfo {author}
  {\bibfnamefont {B.}~\bibnamefont {Knuffman}}, \bibinfo {author}
  {\bibfnamefont {P.~R.}\ \bibnamefont {Berman}}, \ and\ \bibinfo {author}
  {\bibfnamefont {G.}~\bibnamefont {Raithel}},\ }\href {\doibase
  10.1103/PhysRevA.72.023405} {\bibfield  {journal} {\bibinfo  {journal} {Phys.
  Rev. A}\ }\textbf {\bibinfo {volume} {72}},\ \bibinfo {pages} {023405}
  (\bibinfo {year} {2005})}\BibitemShut {NoStop}%
\bibitem [{\citenamefont {Saffman}\ \emph {et~al.}(2010)\citenamefont
  {Saffman}, \citenamefont {Walker},\ and\ \citenamefont
  {M\o{}lmer}}]{RevModPhys.82.2313}%
  \BibitemOpen
  \bibfield  {author} {\bibinfo {author} {\bibfnamefont {M.}~\bibnamefont
  {Saffman}}, \bibinfo {author} {\bibfnamefont {T.~G.}\ \bibnamefont {Walker}},
  \ and\ \bibinfo {author} {\bibfnamefont {K.}~\bibnamefont {M\o{}lmer}},\
  }\href {\doibase 10.1103/RevModPhys.82.2313} {\bibfield  {journal} {\bibinfo
  {journal} {Rev. Mod. Phys.}\ }\textbf {\bibinfo {volume} {82}},\ \bibinfo
  {pages} {2313} (\bibinfo {year} {2010})}\BibitemShut {NoStop}%
\bibitem [{\citenamefont {Timoney}\ \emph {et~al.}(2011)\citenamefont
  {Timoney}, \citenamefont {Baumgart}, \citenamefont {Johanning}, \citenamefont
  {Var{\'o}n}, \citenamefont {Plenio}, \citenamefont {Retzker},\ and\
  \citenamefont {Wunderlich}}]{timoney2011quantum}%
  \BibitemOpen
  \bibfield  {author} {\bibinfo {author} {\bibfnamefont {N.}~\bibnamefont
  {Timoney}}, \bibinfo {author} {\bibfnamefont {I.}~\bibnamefont {Baumgart}},
  \bibinfo {author} {\bibfnamefont {M.}~\bibnamefont {Johanning}}, \bibinfo
  {author} {\bibfnamefont {A.}~\bibnamefont {Var{\'o}n}}, \bibinfo {author}
  {\bibfnamefont {M.~B.}\ \bibnamefont {Plenio}}, \bibinfo {author}
  {\bibfnamefont {A.}~\bibnamefont {Retzker}}, \ and\ \bibinfo {author}
  {\bibfnamefont {C.}~\bibnamefont {Wunderlich}},\ }\href@noop {} {\bibfield
  {journal} {\bibinfo  {journal} {Nature}\ }\textbf {\bibinfo {volume} {476}},\
  \bibinfo {pages} {185} (\bibinfo {year} {2011})}\BibitemShut {NoStop}%
\bibitem [{\citenamefont {Webster}\ \emph {et~al.}(2013)\citenamefont
  {Webster}, \citenamefont {Weidt}, \citenamefont {Lake}, \citenamefont
  {McLoughlin},\ and\ \citenamefont {Hensinger}}]{PhysRevLett.111.140501}%
  \BibitemOpen
  \bibfield  {author} {\bibinfo {author} {\bibfnamefont {S.~C.}\ \bibnamefont
  {Webster}}, \bibinfo {author} {\bibfnamefont {S.}~\bibnamefont {Weidt}},
  \bibinfo {author} {\bibfnamefont {K.}~\bibnamefont {Lake}}, \bibinfo {author}
  {\bibfnamefont {J.~J.}\ \bibnamefont {McLoughlin}}, \ and\ \bibinfo {author}
  {\bibfnamefont {W.~K.}\ \bibnamefont {Hensinger}},\ }\href {\doibase
  10.1103/PhysRevLett.111.140501} {\bibfield  {journal} {\bibinfo  {journal}
  {Phys. Rev. Lett.}\ }\textbf {\bibinfo {volume} {111}},\ \bibinfo {pages}
  {140501} (\bibinfo {year} {2013})}\BibitemShut {NoStop}%
\bibitem [{\citenamefont {Koh}\ \emph {et~al.}(2013)\citenamefont {Koh},
  \citenamefont {Coppersmith},\ and\ \citenamefont {Friesen}}]{koh2013high}%
  \BibitemOpen
  \bibfield  {author} {\bibinfo {author} {\bibfnamefont {T.~S.}\ \bibnamefont
  {Koh}}, \bibinfo {author} {\bibfnamefont {S.}~\bibnamefont {Coppersmith}}, \
  and\ \bibinfo {author} {\bibfnamefont {M.}~\bibnamefont {Friesen}},\
  }\href@noop {} {\bibfield  {journal} {\bibinfo  {journal} {Proceedings of the
  National Academy of Sciences}\ }\textbf {\bibinfo {volume} {110}},\ \bibinfo
  {pages} {19695} (\bibinfo {year} {2013})}\BibitemShut {NoStop}%
\bibitem [{\citenamefont {You}\ and\ \citenamefont
  {Nori}(2011)}]{you2011atomic}%
  \BibitemOpen
  \bibfield  {author} {\bibinfo {author} {\bibfnamefont {J.-Q.}\ \bibnamefont
  {You}}\ and\ \bibinfo {author} {\bibfnamefont {F.}~\bibnamefont {Nori}},\
  }\href@noop {} {\bibfield  {journal} {\bibinfo  {journal} {Nature}\ }\textbf
  {\bibinfo {volume} {474}},\ \bibinfo {pages} {589} (\bibinfo {year}
  {2011})}\BibitemShut {NoStop}%
\bibitem [{\citenamefont {Garcia}\ \emph {et~al.}(2020)\citenamefont {Garcia},
  \citenamefont {Ferri}, \citenamefont {Reichel},\ and\ \citenamefont
  {Long}}]{garcia2020overlapping}%
  \BibitemOpen
  \bibfield  {author} {\bibinfo {author} {\bibfnamefont {S.}~\bibnamefont
  {Garcia}}, \bibinfo {author} {\bibfnamefont {F.}~\bibnamefont {Ferri}},
  \bibinfo {author} {\bibfnamefont {J.}~\bibnamefont {Reichel}}, \ and\
  \bibinfo {author} {\bibfnamefont {R.}~\bibnamefont {Long}},\ }\href@noop {}
  {\bibfield  {journal} {\bibinfo  {journal} {Optics Express}\ }\textbf
  {\bibinfo {volume} {28}},\ \bibinfo {pages} {15515} (\bibinfo {year}
  {2020})}\BibitemShut {NoStop}%
\bibitem [{\citenamefont {Jessen}\ \emph {et~al.}(2001)\citenamefont {Jessen},
  \citenamefont {Haycock}, \citenamefont {Klose}, \citenamefont {Smith},
  \citenamefont {Deutsch},\ and\ \citenamefont {Brennen}}]{jessen2001quantum}%
  \BibitemOpen
  \bibfield  {author} {\bibinfo {author} {\bibfnamefont {P.}~\bibnamefont
  {Jessen}}, \bibinfo {author} {\bibfnamefont {D.}~\bibnamefont {Haycock}},
  \bibinfo {author} {\bibfnamefont {G.}~\bibnamefont {Klose}}, \bibinfo
  {author} {\bibfnamefont {G.}~\bibnamefont {Smith}}, \bibinfo {author}
  {\bibfnamefont {I.}~\bibnamefont {Deutsch}}, \ and\ \bibinfo {author}
  {\bibfnamefont {G.}~\bibnamefont {Brennen}},\ }\href@noop {} {\bibfield
  {journal} {\bibinfo  {journal} {Quantum Information \& Computation}\ }\textbf
  {\bibinfo {volume} {1}},\ \bibinfo {pages} {20} (\bibinfo {year}
  {2001})}\BibitemShut {NoStop}%
\bibitem [{\citenamefont {Zhou}\ \emph {et~al.}(2000)\citenamefont {Zhou},
  \citenamefont {Leung},\ and\ \citenamefont {Chuang}}]{ZLC00}%
  \BibitemOpen
  \bibfield  {author} {\bibinfo {author} {\bibfnamefont {X.}~\bibnamefont
  {Zhou}}, \bibinfo {author} {\bibfnamefont {D.~W.}\ \bibnamefont {Leung}}, \
  and\ \bibinfo {author} {\bibfnamefont {I.~L.}\ \bibnamefont {Chuang}},\
  }\href {\doibase 10.1103/PhysRevA.62.052316} {\bibfield  {journal} {\bibinfo
  {journal} {Phys. Rev. A}\ }\textbf {\bibinfo {volume} {62}},\ \bibinfo
  {pages} {052316} (\bibinfo {year} {2000})}\BibitemShut {NoStop}%
\bibitem [{\citenamefont {Campbell}\ \emph {et~al.}(2017)\citenamefont
  {Campbell}, \citenamefont {Terhal},\ and\ \citenamefont
  {Vuillot}}]{campbell2017roads}%
  \BibitemOpen
  \bibfield  {author} {\bibinfo {author} {\bibfnamefont {E.~T.}\ \bibnamefont
  {Campbell}}, \bibinfo {author} {\bibfnamefont {B.~M.}\ \bibnamefont
  {Terhal}}, \ and\ \bibinfo {author} {\bibfnamefont {C.}~\bibnamefont
  {Vuillot}},\ }\href@noop {} {\bibfield  {journal} {\bibinfo  {journal}
  {Nature}\ }\textbf {\bibinfo {volume} {549}},\ \bibinfo {pages} {172}
  (\bibinfo {year} {2017})}\BibitemShut {NoStop}%
\bibitem [{\citenamefont {M\o{}ller}\ \emph {et~al.}(2007)\citenamefont
  {M\o{}ller}, \citenamefont {Madsen},\ and\ \citenamefont
  {M\o{}lmer}}]{PhysRevA.75.062302}%
  \BibitemOpen
  \bibfield  {author} {\bibinfo {author} {\bibfnamefont {D.}~\bibnamefont
  {M\o{}ller}}, \bibinfo {author} {\bibfnamefont {L.~B.}\ \bibnamefont
  {Madsen}}, \ and\ \bibinfo {author} {\bibfnamefont {K.}~\bibnamefont
  {M\o{}lmer}},\ }\href {\doibase 10.1103/PhysRevA.75.062302} {\bibfield
  {journal} {\bibinfo  {journal} {Phys. Rev. A}\ }\textbf {\bibinfo {volume}
  {75}},\ \bibinfo {pages} {062302} (\bibinfo {year} {2007})}\BibitemShut
  {NoStop}%
\bibitem [{\citenamefont {Roffe}(2019)}]{roffe2019quantum}%
  \BibitemOpen
  \bibfield  {author} {\bibinfo {author} {\bibfnamefont {J.}~\bibnamefont
  {Roffe}},\ }\href@noop {} {\bibfield  {journal} {\bibinfo  {journal}
  {Contemporary Physics}\ }\textbf {\bibinfo {volume} {60}},\ \bibinfo {pages}
  {226} (\bibinfo {year} {2019})}\BibitemShut {NoStop}%
\bibitem [{\citenamefont {Hoeffding}(1963)}]{Hoe63}%
  \BibitemOpen
  \bibfield  {author} {\bibinfo {author} {\bibfnamefont {W.}~\bibnamefont
  {Hoeffding}},\ }\href@noop {} {\bibfield  {journal} {\bibinfo  {journal}
  {Amer. Stat. Assoc, J.}\ }\textbf {\bibinfo {volume} {58}},\ \bibinfo {pages}
  {13} (\bibinfo {year} {1963})}\BibitemShut {NoStop}%
\bibitem [{\citenamefont {Chernoff}\ \emph {et~al.}(1952)\citenamefont
  {Chernoff} \emph {et~al.}}]{chernoff1952measure}%
  \BibitemOpen
  \bibfield  {author} {\bibinfo {author} {\bibfnamefont {H.}~\bibnamefont
  {Chernoff}} \emph {et~al.},\ }\href@noop {} {\bibfield  {journal} {\bibinfo
  {journal} {The Annals of Mathematical Statistics}\ }\textbf {\bibinfo
  {volume} {23}},\ \bibinfo {pages} {493} (\bibinfo {year} {1952})}\BibitemShut
  {NoStop}%
\bibitem [{\citenamefont {Okamoto}(1959)}]{okamoto1959some}%
  \BibitemOpen
  \bibfield  {author} {\bibinfo {author} {\bibfnamefont {M.}~\bibnamefont
  {Okamoto}},\ }\href@noop {} {\bibfield  {journal} {\bibinfo  {journal}
  {Annals of the institute of Statistical Mathematics}\ }\textbf {\bibinfo
  {volume} {10}},\ \bibinfo {pages} {29} (\bibinfo {year} {1959})}\BibitemShut
  {NoStop}%
\bibitem [{\citenamefont {Feng}\ and\ \citenamefont {Ma}(2004)}]{FeM04}%
  \BibitemOpen
  \bibfield  {author} {\bibinfo {author} {\bibfnamefont {K.}~\bibnamefont
  {Feng}}\ and\ \bibinfo {author} {\bibfnamefont {Z.}~\bibnamefont {Ma}},\
  }\href@noop {} {\bibfield  {journal} {\bibinfo  {journal} {IEEE Transactions
  on Information Theory}\ }\textbf {\bibinfo {volume} {50}},\ \bibinfo {pages}
  {3323} (\bibinfo {year} {2004})}\BibitemShut {NoStop}%
\bibitem [{\citenamefont {Ma}(2008)}]{MaY08}%
  \BibitemOpen
  \bibfield  {author} {\bibinfo {author} {\bibfnamefont {Y.}~\bibnamefont
  {Ma}},\ }\href@noop {} {\bibfield  {journal} {\bibinfo  {journal} {Journal of
  Mathematical Analysis and Applications}\ }\textbf {\bibinfo {volume} {340}},\
  \bibinfo {pages} {550} (\bibinfo {year} {2008})}\BibitemShut {NoStop}%
\bibitem [{\citenamefont {Jin}\ and\ \citenamefont {Xing}(2011)}]{JiX11}%
  \BibitemOpen
  \bibfield  {author} {\bibinfo {author} {\bibfnamefont {L.}~\bibnamefont
  {Jin}}\ and\ \bibinfo {author} {\bibfnamefont {C.}~\bibnamefont {Xing}},\
  }in\ \href {\doibase 10.1109/ISIT.2011.6034167} {\emph {\bibinfo {booktitle}
  {IEEE International Symposium on Information Theory Proceedings (ISIT)}}}\
  (\bibinfo {year} {2011})\ pp.\ \bibinfo {pages} {455--458}\BibitemShut
  {NoStop}%
\bibitem [{\citenamefont {{Ouyang}}(2014)}]{ouyang2014concatenated}%
  \BibitemOpen
  \bibfield  {author} {\bibinfo {author} {\bibfnamefont {Y.}~\bibnamefont
  {{Ouyang}}},\ }\href {\doibase 10.1109/TIT.2014.2313577} {\bibfield
  {journal} {\bibinfo  {journal} {IEEE Transactions on Information Theory}\
  }\textbf {\bibinfo {volume} {60}},\ \bibinfo {pages} {3117} (\bibinfo {year}
  {2014})}\BibitemShut {NoStop}%
\bibitem [{\citenamefont {Leung}\ \emph {et~al.}(1997)\citenamefont {Leung},
  \citenamefont {Nielsen}, \citenamefont {Chuang},\ and\ \citenamefont
  {Yamamoto}}]{LNCY97}%
  \BibitemOpen
  \bibfield  {author} {\bibinfo {author} {\bibfnamefont {D.~W.}\ \bibnamefont
  {Leung}}, \bibinfo {author} {\bibfnamefont {M.~A.}\ \bibnamefont {Nielsen}},
  \bibinfo {author} {\bibfnamefont {I.~L.}\ \bibnamefont {Chuang}}, \ and\
  \bibinfo {author} {\bibfnamefont {Y.}~\bibnamefont {Yamamoto}},\ }\href
  {\doibase 10.1103/PhysRevA.56.2567} {\bibfield  {journal} {\bibinfo
  {journal} {Phys. Rev. A}\ }\textbf {\bibinfo {volume} {56}},\ \bibinfo
  {pages} {2567} (\bibinfo {year} {1997})}\BibitemShut {NoStop}%
\bibitem [{\citenamefont {Ouyang}(2014)}]{ouyang2014permutation}%
  \BibitemOpen
  \bibfield  {author} {\bibinfo {author} {\bibfnamefont {Y.}~\bibnamefont
  {Ouyang}},\ }\href {\doibase 10.1103/PhysRevA.90.062317} {\bibfield
  {journal} {\bibinfo  {journal} {Phys. Rev. A}\ }\textbf {\bibinfo {volume}
  {90}},\ \bibinfo {pages} {062317} (\bibinfo {year} {2014})},\ \Eprint
  {http://arxiv.org/abs/1302.3247} {1302.3247} \BibitemShut {NoStop}%
\bibitem [{\citenamefont {Tuckett}\ \emph {et~al.}(2018)\citenamefont
  {Tuckett}, \citenamefont {Bartlett},\ and\ \citenamefont
  {Flammia}}]{PhysRevLett.120.050505}%
  \BibitemOpen
  \bibfield  {author} {\bibinfo {author} {\bibfnamefont {D.~K.}\ \bibnamefont
  {Tuckett}}, \bibinfo {author} {\bibfnamefont {S.~D.}\ \bibnamefont
  {Bartlett}}, \ and\ \bibinfo {author} {\bibfnamefont {S.~T.}\ \bibnamefont
  {Flammia}},\ }\href {\doibase 10.1103/PhysRevLett.120.050505} {\bibfield
  {journal} {\bibinfo  {journal} {Phys. Rev. Lett.}\ }\textbf {\bibinfo
  {volume} {120}},\ \bibinfo {pages} {050505} (\bibinfo {year}
  {2018})}\BibitemShut {NoStop}%
\bibitem [{\citenamefont {Ashikhmin}\ \emph {et~al.}(2001)\citenamefont
  {Ashikhmin}, \citenamefont {Litsyn},\ and\ \citenamefont {Tsfasman}}]{ALT01}%
  \BibitemOpen
  \bibfield  {author} {\bibinfo {author} {\bibfnamefont {A.}~\bibnamefont
  {Ashikhmin}}, \bibinfo {author} {\bibfnamefont {S.}~\bibnamefont {Litsyn}}, \
  and\ \bibinfo {author} {\bibfnamefont {M.~A.}\ \bibnamefont {Tsfasman}},\
  }\href {\doibase 10.1103/PhysRevA.63.032311} {\bibfield  {journal} {\bibinfo
  {journal} {Phys. Rev. A}\ }\textbf {\bibinfo {volume} {63}},\ \bibinfo
  {pages} {32311} (\bibinfo {year} {2001})}\BibitemShut {NoStop}%
\bibitem [{\citenamefont {Chen}\ \emph {et~al.}(2001)\citenamefont {Chen},
  \citenamefont {Ling},\ and\ \citenamefont {Xing}}]{CLX01}%
  \BibitemOpen
  \bibfield  {author} {\bibinfo {author} {\bibfnamefont {H.}~\bibnamefont
  {Chen}}, \bibinfo {author} {\bibfnamefont {S.}~\bibnamefont {Ling}}, \ and\
  \bibinfo {author} {\bibfnamefont {C.}~\bibnamefont {Xing}},\ }\href {\doibase
  10.1109/18.930941} {\bibfield  {journal} {\bibinfo  {journal} {IEEE
  Transactions on Information Theory}\ }\textbf {\bibinfo {volume} {47}},\
  \bibinfo {pages} {2055} (\bibinfo {year} {2001})}\BibitemShut {NoStop}%
\bibitem [{\citenamefont {Li}\ \emph {et~al.}(2009)\citenamefont {Li},
  \citenamefont {Xing},\ and\ \citenamefont {Wang}}]{LXW09}%
  \BibitemOpen
  \bibfield  {author} {\bibinfo {author} {\bibfnamefont {Z.}~\bibnamefont
  {Li}}, \bibinfo {author} {\bibfnamefont {L.}~\bibnamefont {Xing}}, \ and\
  \bibinfo {author} {\bibfnamefont {X.}~\bibnamefont {Wang}},\ }\href {\doibase
  10.1109/TIT.2009.2023715} {\bibfield  {journal} {\bibinfo  {journal} {IEEE
  Transactions on Information Theory}\ }\textbf {\bibinfo {volume} {55}},\
  \bibinfo {pages} {3821} (\bibinfo {year} {2009})}\BibitemShut {NoStop}%
\bibitem [{\citenamefont {Darmawan}\ \emph {et~al.}(2021)\citenamefont
  {Darmawan}, \citenamefont {Brown}, \citenamefont {Grimsmo}, \citenamefont
  {Tuckett},\ and\ \citenamefont {Puri}}]{PRXQuantum.2.030345}%
  \BibitemOpen
  \bibfield  {author} {\bibinfo {author} {\bibfnamefont {A.~S.}\ \bibnamefont
  {Darmawan}}, \bibinfo {author} {\bibfnamefont {B.~J.}\ \bibnamefont {Brown}},
  \bibinfo {author} {\bibfnamefont {A.~L.}\ \bibnamefont {Grimsmo}}, \bibinfo
  {author} {\bibfnamefont {D.~K.}\ \bibnamefont {Tuckett}}, \ and\ \bibinfo
  {author} {\bibfnamefont {S.}~\bibnamefont {Puri}},\ }\href {\doibase
  10.1103/PRXQuantum.2.030345} {\bibfield  {journal} {\bibinfo  {journal} {PRX
  Quantum}\ }\textbf {\bibinfo {volume} {2}},\ \bibinfo {pages} {030345}
  (\bibinfo {year} {2021})}\BibitemShut {NoStop}%
\bibitem [{\citenamefont {Fowler}\ \emph {et~al.}(2012)\citenamefont {Fowler},
  \citenamefont {Mariantoni}, \citenamefont {Martinis},\ and\ \citenamefont
  {Cleland}}]{Cleland12}%
  \BibitemOpen
  \bibfield  {author} {\bibinfo {author} {\bibfnamefont {A.~G.}\ \bibnamefont
  {Fowler}}, \bibinfo {author} {\bibfnamefont {M.}~\bibnamefont {Mariantoni}},
  \bibinfo {author} {\bibfnamefont {J.~M.}\ \bibnamefont {Martinis}}, \ and\
  \bibinfo {author} {\bibfnamefont {A.~N.}\ \bibnamefont {Cleland}},\ }\href
  {\doibase 10.1103/PhysRevA.86.032324} {\bibfield  {journal} {\bibinfo
  {journal} {Phys. Rev. A}\ }\textbf {\bibinfo {volume} {86}},\ \bibinfo
  {pages} {032324} (\bibinfo {year} {2012})}\BibitemShut {NoStop}%
\bibitem [{\citenamefont {Wu}\ \emph {et~al.}(2022)\citenamefont {Wu},
  \citenamefont {Kolkowitz}, \citenamefont {Puri},\ and\ \citenamefont
  {Thompson}}]{wu2022erasure}%
  \BibitemOpen
  \bibfield  {author} {\bibinfo {author} {\bibfnamefont {Y.}~\bibnamefont
  {Wu}}, \bibinfo {author} {\bibfnamefont {S.}~\bibnamefont {Kolkowitz}},
  \bibinfo {author} {\bibfnamefont {S.}~\bibnamefont {Puri}}, \ and\ \bibinfo
  {author} {\bibfnamefont {J.~D.}\ \bibnamefont {Thompson}},\ }\href@noop {}
  {\bibfield  {journal} {\bibinfo  {journal} {arXiv preprint arXiv:2201.03540}\
  } (\bibinfo {year} {2022})}\BibitemShut {NoStop}%
\bibitem [{\citenamefont {Panteleev}\ and\ \citenamefont
  {Kalachev}(2022)}]{panteleev2022asymptotically}%
  \BibitemOpen
  \bibfield  {author} {\bibinfo {author} {\bibfnamefont {P.}~\bibnamefont
  {Panteleev}}\ and\ \bibinfo {author} {\bibfnamefont {G.}~\bibnamefont
  {Kalachev}},\ }in\ \href@noop {} {\emph {\bibinfo {booktitle} {Proceedings of
  the 54th Annual ACM SIGACT Symposium on Theory of Computing}}}\ (\bibinfo
  {year} {2022})\ pp.\ \bibinfo {pages} {375--388}\BibitemShut {NoStop}%
\bibitem [{\citenamefont {Ouyang}(2021)}]{ouyang2021permutation}%
  \BibitemOpen
  \bibfield  {author} {\bibinfo {author} {\bibfnamefont {Y.}~\bibnamefont
  {Ouyang}},\ }in\ \href {\doibase 10.1109/ISIT45174.2021.9518078} {\emph
  {\bibinfo {booktitle} {2021 IEEE International Symposium on Information
  Theory (ISIT)}}}\ (\bibinfo {year} {2021})\ pp.\ \bibinfo {pages}
  {1499--1503}\BibitemShut {NoStop}%
\bibitem [{\citenamefont {Shibayama}\ and\ \citenamefont
  {Hagiwara}(2021)}]{shibayama2021permutation}%
  \BibitemOpen
  \bibfield  {author} {\bibinfo {author} {\bibfnamefont {T.}~\bibnamefont
  {Shibayama}}\ and\ \bibinfo {author} {\bibfnamefont {M.}~\bibnamefont
  {Hagiwara}},\ }\bibfield  {booktitle} {\emph {\bibinfo {booktitle} {2021 IEEE
  International Symposium on Information Theory (ISIT)}},\ }\href@noop {} {\ ,\
  \bibinfo {pages} {1493} (\bibinfo {year} {2021})}\BibitemShut {NoStop}%
\bibitem [{\citenamefont {Nakayama}\ and\ \citenamefont
  {Hagiwara}(2020)}]{nakayama2020first}%
  \BibitemOpen
  \bibfield  {author} {\bibinfo {author} {\bibfnamefont {A.}~\bibnamefont
  {Nakayama}}\ and\ \bibinfo {author} {\bibfnamefont {M.}~\bibnamefont
  {Hagiwara}},\ }\href@noop {} {\bibfield  {journal} {\bibinfo  {journal}
  {IEICE Communications Express}\ }\textbf {\bibinfo {volume} {9}},\ \bibinfo
  {pages} {100} (\bibinfo {year} {2020})}\BibitemShut {NoStop}%
\bibitem [{\citenamefont {Shibayama}\ and\ \citenamefont
  {Ouyang}(2021)}]{shibayama2021equivalence}%
  \BibitemOpen
  \bibfield  {author} {\bibinfo {author} {\bibfnamefont {T.}~\bibnamefont
  {Shibayama}}\ and\ \bibinfo {author} {\bibfnamefont {Y.}~\bibnamefont
  {Ouyang}},\ }in\ \href@noop {} {\emph {\bibinfo {booktitle} {2021 IEEE
  Information Theory Workshop (ITW)}}}\ (\bibinfo {organization} {IEEE},\
  \bibinfo {year} {2021})\ pp.\ \bibinfo {pages} {1--6}\BibitemShut {NoStop}%
\bibitem [{\citenamefont {Baireuther}\ \emph {et~al.}(2019)\citenamefont
  {Baireuther}, \citenamefont {Caio}, \citenamefont {Criger}, \citenamefont
  {Beenakker},\ and\ \citenamefont {O’Brien}}]{baireuther2019neural}%
  \BibitemOpen
  \bibfield  {author} {\bibinfo {author} {\bibfnamefont {P.}~\bibnamefont
  {Baireuther}}, \bibinfo {author} {\bibfnamefont {M.~D.}\ \bibnamefont
  {Caio}}, \bibinfo {author} {\bibfnamefont {B.}~\bibnamefont {Criger}},
  \bibinfo {author} {\bibfnamefont {C.~W.}\ \bibnamefont {Beenakker}}, \ and\
  \bibinfo {author} {\bibfnamefont {T.~E.}\ \bibnamefont {O’Brien}},\
  }\href@noop {} {\bibfield  {journal} {\bibinfo  {journal} {New Journal of
  Physics}\ }\textbf {\bibinfo {volume} {21}},\ \bibinfo {pages} {013003}
  (\bibinfo {year} {2019})}\BibitemShut {NoStop}%
\bibitem [{\citenamefont {Chao}\ and\ \citenamefont
  {Reichardt}(2020)}]{Chao2020.anyflag}%
  \BibitemOpen
  \bibfield  {author} {\bibinfo {author} {\bibfnamefont {R.}~\bibnamefont
  {Chao}}\ and\ \bibinfo {author} {\bibfnamefont {B.~W.}\ \bibnamefont
  {Reichardt}},\ }\href {\doibase 10.1103/PRXQuantum.1.010302} {\bibfield
  {journal} {\bibinfo  {journal} {PRX Quantum}\ }\textbf {\bibinfo {volume}
  {1}},\ \bibinfo {pages} {010302} (\bibinfo {year} {2020})}\BibitemShut
  {NoStop}%
\bibitem [{\citenamefont {Goldberg}\ and\ \citenamefont
  {James}(2019)}]{PhysRevA.100.042332}%
  \BibitemOpen
  \bibfield  {author} {\bibinfo {author} {\bibfnamefont {A.~Z.}\ \bibnamefont
  {Goldberg}}\ and\ \bibinfo {author} {\bibfnamefont {D.~F.~V.}\ \bibnamefont
  {James}},\ }\href {\doibase 10.1103/PhysRevA.100.042332} {\bibfield
  {journal} {\bibinfo  {journal} {Phys. Rev. A}\ }\textbf {\bibinfo {volume}
  {100}},\ \bibinfo {pages} {042332} (\bibinfo {year} {2019})}\BibitemShut
  {NoStop}%
\bibitem [{\citenamefont {Tsang}(2019)}]{tsang2019resolving}%
  \BibitemOpen
  \bibfield  {author} {\bibinfo {author} {\bibfnamefont {M.}~\bibnamefont
  {Tsang}},\ }\href {\doibase 10.1080/00107514.2020.1736375} {\bibfield
  {journal} {\bibinfo  {journal} {Contemporary Physics}\ }\textbf {\bibinfo
  {volume} {60}},\ \bibinfo {pages} {279} (\bibinfo {year} {2019})}\BibitemShut
  {NoStop}%
\bibitem [{\citenamefont {Nair}\ and\ \citenamefont
  {Tsang}(2016)}]{PhysRevLett.117.190801}%
  \BibitemOpen
  \bibfield  {author} {\bibinfo {author} {\bibfnamefont {R.}~\bibnamefont
  {Nair}}\ and\ \bibinfo {author} {\bibfnamefont {M.}~\bibnamefont {Tsang}},\
  }\href {\doibase 10.1103/PhysRevLett.117.190801} {\bibfield  {journal}
  {\bibinfo  {journal} {Phys. Rev. Lett.}\ }\textbf {\bibinfo {volume} {117}},\
  \bibinfo {pages} {190801} (\bibinfo {year} {2016})}\BibitemShut {NoStop}%
\bibitem [{\citenamefont {Huang}\ and\ \citenamefont
  {Lupo}(2021)}]{PhysRevLett.127.130502}%
  \BibitemOpen
  \bibfield  {author} {\bibinfo {author} {\bibfnamefont {Z.}~\bibnamefont
  {Huang}}\ and\ \bibinfo {author} {\bibfnamefont {C.}~\bibnamefont {Lupo}},\
  }\href {\doibase 10.1103/PhysRevLett.127.130502} {\bibfield  {journal}
  {\bibinfo  {journal} {Phys. Rev. Lett.}\ }\textbf {\bibinfo {volume} {127}},\
  \bibinfo {pages} {130502} (\bibinfo {year} {2021})}\BibitemShut {NoStop}%
\bibitem [{\citenamefont {Huang}\ \emph {et~al.}(2021)\citenamefont {Huang},
  \citenamefont {Lupo},\ and\ \citenamefont {Kok}}]{PRXQuantum.2.030303}%
  \BibitemOpen
  \bibfield  {author} {\bibinfo {author} {\bibfnamefont {Z.}~\bibnamefont
  {Huang}}, \bibinfo {author} {\bibfnamefont {C.}~\bibnamefont {Lupo}}, \ and\
  \bibinfo {author} {\bibfnamefont {P.}~\bibnamefont {Kok}},\ }\href {\doibase
  10.1103/PRXQuantum.2.030303} {\bibfield  {journal} {\bibinfo  {journal} {PRX
  Quantum}\ }\textbf {\bibinfo {volume} {2}},\ \bibinfo {pages} {030303}
  (\bibinfo {year} {2021})}\BibitemShut {NoStop}%
\bibitem [{\citenamefont {Backlund}\ \emph {et~al.}(2018)\citenamefont
  {Backlund}, \citenamefont {Shechtman},\ and\ \citenamefont
  {Walsworth}}]{PhysRevLett.121.023904}%
  \BibitemOpen
  \bibfield  {author} {\bibinfo {author} {\bibfnamefont {M.~P.}\ \bibnamefont
  {Backlund}}, \bibinfo {author} {\bibfnamefont {Y.}~\bibnamefont {Shechtman}},
  \ and\ \bibinfo {author} {\bibfnamefont {R.~L.}\ \bibnamefont {Walsworth}},\
  }\href {\doibase 10.1103/PhysRevLett.121.023904} {\bibfield  {journal}
  {\bibinfo  {journal} {Phys. Rev. Lett.}\ }\textbf {\bibinfo {volume} {121}},\
  \bibinfo {pages} {023904} (\bibinfo {year} {2018})}\BibitemShut {NoStop}%
\bibitem [{\citenamefont {Yu}\ and\ \citenamefont
  {Prasad}(2018)}]{PhysRevLett.121.180504}%
  \BibitemOpen
  \bibfield  {author} {\bibinfo {author} {\bibfnamefont {Z.}~\bibnamefont
  {Yu}}\ and\ \bibinfo {author} {\bibfnamefont {S.}~\bibnamefont {Prasad}},\
  }\href {\doibase 10.1103/PhysRevLett.121.180504} {\bibfield  {journal}
  {\bibinfo  {journal} {Phys. Rev. Lett.}\ }\textbf {\bibinfo {volume} {121}},\
  \bibinfo {pages} {180504} (\bibinfo {year} {2018})}\BibitemShut {NoStop}%
\bibitem [{\citenamefont {Napoli}\ \emph {et~al.}(2019)\citenamefont {Napoli},
  \citenamefont {Piano}, \citenamefont {Leach}, \citenamefont {Adesso},\ and\
  \citenamefont {Tufarelli}}]{PhysRevLett.122.140505}%
  \BibitemOpen
  \bibfield  {author} {\bibinfo {author} {\bibfnamefont {C.}~\bibnamefont
  {Napoli}}, \bibinfo {author} {\bibfnamefont {S.}~\bibnamefont {Piano}},
  \bibinfo {author} {\bibfnamefont {R.}~\bibnamefont {Leach}}, \bibinfo
  {author} {\bibfnamefont {G.}~\bibnamefont {Adesso}}, \ and\ \bibinfo {author}
  {\bibfnamefont {T.}~\bibnamefont {Tufarelli}},\ }\href {\doibase
  10.1103/PhysRevLett.122.140505} {\bibfield  {journal} {\bibinfo  {journal}
  {Phys. Rev. Lett.}\ }\textbf {\bibinfo {volume} {122}},\ \bibinfo {pages}
  {140505} (\bibinfo {year} {2019})}\BibitemShut {NoStop}%
\bibitem [{\citenamefont {Tham}\ \emph {et~al.}(2017)\citenamefont {Tham},
  \citenamefont {Ferretti},\ and\ \citenamefont
  {Steinberg}}]{PhysRevLett.118.070801}%
  \BibitemOpen
  \bibfield  {author} {\bibinfo {author} {\bibfnamefont {W.-K.}\ \bibnamefont
  {Tham}}, \bibinfo {author} {\bibfnamefont {H.}~\bibnamefont {Ferretti}}, \
  and\ \bibinfo {author} {\bibfnamefont {A.~M.}\ \bibnamefont {Steinberg}},\
  }\href {\doibase 10.1103/PhysRevLett.118.070801} {\bibfield  {journal}
  {\bibinfo  {journal} {Phys. Rev. Lett.}\ }\textbf {\bibinfo {volume} {118}},\
  \bibinfo {pages} {070801} (\bibinfo {year} {2017})}\BibitemShut {NoStop}%
\bibitem [{\citenamefont {Parniak}\ \emph {et~al.}(2018)\citenamefont
  {Parniak}, \citenamefont {Bor\'owka}, \citenamefont {Boroszko}, \citenamefont
  {Wasilewski}, \citenamefont {Banaszek},\ and\ \citenamefont
  {Demkowicz-Dobrza\ifmmode~\acute{n}\else
  \'{n}\fi{}ski}}]{PhysRevLett.121.250503}%
  \BibitemOpen
  \bibfield  {author} {\bibinfo {author} {\bibfnamefont {M.}~\bibnamefont
  {Parniak}}, \bibinfo {author} {\bibfnamefont {S.}~\bibnamefont {Bor\'owka}},
  \bibinfo {author} {\bibfnamefont {K.}~\bibnamefont {Boroszko}}, \bibinfo
  {author} {\bibfnamefont {W.}~\bibnamefont {Wasilewski}}, \bibinfo {author}
  {\bibfnamefont {K.}~\bibnamefont {Banaszek}}, \ and\ \bibinfo {author}
  {\bibfnamefont {R.}~\bibnamefont {Demkowicz-Dobrza\ifmmode~\acute{n}\else
  \'{n}\fi{}ski}},\ }\href {\doibase 10.1103/PhysRevLett.121.250503} {\bibfield
   {journal} {\bibinfo  {journal} {Phys. Rev. Lett.}\ }\textbf {\bibinfo
  {volume} {121}},\ \bibinfo {pages} {250503} (\bibinfo {year}
  {2018})}\BibitemShut {NoStop}%
\bibitem [{\citenamefont {Hassett}\ \emph {et~al.}(2018)\citenamefont
  {Hassett}, \citenamefont {Malhorta}, \citenamefont {Alonso}, \citenamefont
  {Boyd}, \citenamefont {Rafsanjani},\ and\ \citenamefont
  {Vamivakas}}]{hassett2018sub}%
  \BibitemOpen
  \bibfield  {author} {\bibinfo {author} {\bibfnamefont {J.}~\bibnamefont
  {Hassett}}, \bibinfo {author} {\bibfnamefont {T.}~\bibnamefont {Malhorta}},
  \bibinfo {author} {\bibfnamefont {M.}~\bibnamefont {Alonso}}, \bibinfo
  {author} {\bibfnamefont {R.}~\bibnamefont {Boyd}}, \bibinfo {author}
  {\bibfnamefont {S.~H.}\ \bibnamefont {Rafsanjani}}, \ and\ \bibinfo {author}
  {\bibfnamefont {A.}~\bibnamefont {Vamivakas}},\ }in\ \href@noop {} {\emph
  {\bibinfo {booktitle} {Laser Science}}}\ (\bibinfo {organization} {Optical
  Society of America},\ \bibinfo {year} {2018})\ pp.\ \bibinfo {pages}
  {JW4A--124}\BibitemShut {NoStop}%
\bibitem [{\citenamefont {Zhou}\ \emph {et~al.}(2019)\citenamefont {Zhou},
  \citenamefont {Yang}, \citenamefont {Hassett}, \citenamefont {Rafsanjani},
  \citenamefont {Mirhosseini}, \citenamefont {Vamivakas}, \citenamefont
  {Jordan}, \citenamefont {Shi},\ and\ \citenamefont {Boyd}}]{zhou2019quantum}%
  \BibitemOpen
  \bibfield  {author} {\bibinfo {author} {\bibfnamefont {Y.}~\bibnamefont
  {Zhou}}, \bibinfo {author} {\bibfnamefont {J.}~\bibnamefont {Yang}}, \bibinfo
  {author} {\bibfnamefont {J.~D.}\ \bibnamefont {Hassett}}, \bibinfo {author}
  {\bibfnamefont {S.~M.~H.}\ \bibnamefont {Rafsanjani}}, \bibinfo {author}
  {\bibfnamefont {M.}~\bibnamefont {Mirhosseini}}, \bibinfo {author}
  {\bibfnamefont {A.~N.}\ \bibnamefont {Vamivakas}}, \bibinfo {author}
  {\bibfnamefont {A.~N.}\ \bibnamefont {Jordan}}, \bibinfo {author}
  {\bibfnamefont {Z.}~\bibnamefont {Shi}}, \ and\ \bibinfo {author}
  {\bibfnamefont {R.~W.}\ \bibnamefont {Boyd}},\ }\href@noop {} {\bibfield
  {journal} {\bibinfo  {journal} {Optica}\ }\textbf {\bibinfo {volume} {6}},\
  \bibinfo {pages} {534} (\bibinfo {year} {2019})}\BibitemShut {NoStop}%
\bibitem [{\citenamefont {Preskill}(2018)}]{preskill2018quantum}%
  \BibitemOpen
  \bibfield  {author} {\bibinfo {author} {\bibfnamefont {J.}~\bibnamefont
  {Preskill}},\ }\href@noop {} {\bibfield  {journal} {\bibinfo  {journal}
  {Quantum}\ }\textbf {\bibinfo {volume} {2}},\ \bibinfo {pages} {79} (\bibinfo
  {year} {2018})}\BibitemShut {NoStop}%
\end{thebibliography}
\end{document}